\documentclass[pra,twocolumn,floatfix,showpacs]{revtex4-1}
\usepackage{graphicx,bm,amsmath,amssymb,amsfonts}
\usepackage{color}

\begin{document}
\title{
Quantum Degenerate Fermi Gas with Spin-orbit Coupling and Crossed Zeeman Fields
}

\author{
Kangjun Seo, Li Han, and C. A. R. S{\'a} de Melo
}

\affiliation{
School of Physics, Georgia Institute of Technology,
Atlanta, Georgia 30332, USA
}

\date{\today}

\begin{abstract}
We study quantum degenerate ultra-cold Fermi gases in the presence of
artificial spin-orbit coupling and crossed Zeeman fields.
We emphasize the case where parity is violated in the excitation
spectrum and compare it with the simpler situation where
parity is preserved.
We investigate in detail spectroscopic properties such as
the excitation spectrum, the spectral function, momentum
distribution and density of states for the cases where parity is
preserved or violated. Similarly,
we show that thermodynamic properties such as pressure,
chemical potential, entropy, specific heat, isothermal
compressibility and induced spin polarization become anisotropic
as a function of Zeeman field components, when parity is violated.
Lastly, we discuss the effects of interactions and
present results for the pairing temperature as the precursor
for the transition to a superfluid state.  In particular, we find
that the pairing temperature is dramatically reduced in the
weak interaction regime as parity violation gets stronger, and that
the momentum dependence of the order parameter for superfluidity
violates parity when crossed Zeeman fields are present for finite
spin-orbit coupling.
\end{abstract}


\maketitle

%
%

%
\section{Introduction}
\label{sec:introduction}

The study of quantum degenerate fermions has been
in the forefront of research in ultra-cold atoms and molecules
in recent years, where particular attention was paid to the
so-called evolution from BCS to BEC
superfluidity. Bringing ultra-cold fermions into quantum degeneracy
and using Feshbach resonances to tune interactions between colliding
fermions opened the door for the exploration of their superfluid
phases, and their thermodynamic and correlation properties.
This ability to tune interactions and explore the limits
of weak, strong and unitary interactions had impact not only in
cold-atom physics, but also in condensed matter physics
(strongly correlated superconductors), nuclear physics
(superconductivity in quantum chromodynamics)
and astrophysics (superfluidity in neutron stars).

Many advances in cold atoms followed after the development of new
tools for their toolbox. For instance the experimental study of
the evolution from BCS to BEC superfluidity occurred after
appropriate Feshbach resonances for $^6$Li and $^{40}$K were
identified, and used to study the crossover problem for $s$-wave superfluids.
In addition to $s$-wave Feshbach resonances, both $^6$Li and $^{40}$K
also exhibit $p$-wave Feshbach resonances, which could produce p-wave
superfluids in particular in BEC regime if $p$-wave Feshbach molecules were
stable. Unfortunately, such $p$-wave molecules do not live sufficiently long
for the creation of $p$-wave superfluids~\cite{jin-2007, mukaiyama-2008,
vale-2008, zimmermann-2010}.

In addition to the manipulation of interactions, it has been possible
to extract experimentally detailed thermodynamic information of
interacting ultra-cold fermions~\cite{salomon-2010, zwierlein-2012a}
from local density images of trapped atoms with the help
of the local density approximation~\cite{yip-2007, ho-2009}.

A further tool was developed recently through the production of
tunable artificial spin-orbit fields~\cite{spielman-2011}
that were created in $^{87}$Rb, a bosonic isotope of Rubidium,
by using a set of Raman beams that allowed momentum transfer
and mixing of two dressed spin states. In these experiments,
the emergence of spin-orbit fields, controlled by the momentum
transfer from the light fields to the atoms, was connected
to the simultaneous existence of an artificial Zeeman field
controlled by the Raman coupling. In such experiments, the
artificial spin-orbit and Zeeman fields were used to manipulate
the effective interactions between bosons and thus produce
new quantum phases of bosonic matter~\cite{spielman-2011}.
In a subsequent experiment it was directly demonstrated that
the effective interactions between bosons can acquire
higher angular momentum components which are directly controlled by
the artificial spin-orbit and Zeeman fields. The effective
single-atom Hamiltonian created in these experiments is
\begin{equation}
\label{eqn:hamiltonian-single-atom}
{\bf H}_{\rm sa} ({\bf k})
=
\epsilon_{\bf k} {\bf 1} - h_\parallel  \sigma_z
- h_x({\bf k})  \sigma_x - h_y ({\bf k}) \sigma_y,
\end{equation}
when expressed in the dressed state spin
basis $\vert {\bf k}, s \rangle$.
Here,
$
\epsilon_{\bf k}
=
{\bf k}^2/(2m)
$
represents the kinetic energy of an atom with momentum
${\bf k}$ and spin state $s$, $h_\parallel = h_z = \Omega/2$
represents the Zeeman field along the quantization axis $z$
which is directly related to the Raman coupling $\Omega$,
$h_x ({\bf k}) = 0$ and $h_y ({\bf k}) = h_y + vk_x$,
where $h_y = \delta/2$ represents the detuning and
$vk_x$ represents a mixture of equal Rashba~\cite{rashba-1960}
and Dresselhaus~\cite{dresselhaus-1955} terms, which we
label as ERD spin-orbit coupling.

The possibility of studying similar phenomena with ultra-cold fermions
in the presence of spin-orbit coupling and Zeeman
fields~\cite{spielman-2011,sademelo-2011} lead to an explosion of
theoretical research, which focused primarily on the case of Rashba-only
(RO) spin-orbit coupling~\cite{shenoy-2011, chuanwei-2011, zhai-2011,
hu-2011, iskin-2011} which has been studied extensively in condensed matter
physics in the context of non-centro-symmetric
superconductors~\cite{gorkov-2001, yip-2002, sigrist-2004}.
Some suggestions of possible realizations of artificial
RO spin-orbit fields in ultra-cold fermions have appeared
in the literature~\cite{chuanwei-2008, sinova-2009}.
However, in our group, we have focused mostly
on the ERD case~\cite{han-2012,seo-2012a,seo-2011,seo-2012b},
which is simpler to be realized experimentally as it was demonstrated
for interacting bosons~\cite{spielman-2011, sademelo-2011}
in the case of $^{87}$Rb, and more recently
for non-interacting fermions such as $^{40}$K in China~\cite{zhang-2012}
and $^6$Li in the United States~\cite{zwierlein-2012b}.

In our recent work~\cite{han-2012, seo-2012a, seo-2011, seo-2012b},
we emphasized the importance of interactions in producing
novel phases for spin-orbit coupled ultra-cold fermions
in the presence of Zeeman fields, and we analyzed in detail the
ERD case with a longitudinal Zeeman field which is orthogonal
to the spin-orbit field~\cite{seo-2012a, seo-2011, seo-2012b}.
In this paper, we generalize our previous analysis to include an
additional Zeeman field $h_y$ representing the detuning $\delta$,
which is perpendicular to
longitudinal Zeeman field $h_\parallel = h_z$ representing the
Raman coupling $\Omega$. In this case, the simultaneous presence of $h_y$
and the ERD spin-orbit field $v k_x$ along the $y$ axis
leads to the loss of parity for the eigenvalues of $H_{\rm sa} ({\bf k})$
defined in Eq.~(\ref{eqn:hamiltonian-single-atom}) above.
This violation of parity in the energy spectrum leads to many detectable
parity-violating properties in spectroscopic quantities such as
the spectral function and momentum distribution, which are explicitly
momentum-dependent, or in momentum-integrated quantities like
the density of states when viewed as a function of
Zeeman fields $h_y$ and $h_y$. Furthermore, we demonstrate that
this parity violation can also be seen and quantified in
thermodynamic quantities such as the pressure,
entropy, specific heat, chemical potential, compressibility and
spin-polarization, which are discussed in detail for non-interacting
ultra-cold fermions in the presence of ERD spin-orbit and crossed
Zeeman fields $h_y$ and $h_z$. Furthermore, we also discuss
the effects of parity violation when attractive $s$-wave interactions
are present, and we pay particular attention to the pairing temperature
and to the absence of parity in the order parameter tensor describing the
superfluid order.

The remainder of the paper is as follows.
In section~\ref{sec:magnetic-hamiltonian}, we describe the magnetic
hamiltonian in its general form, and particularize it to any linear
combination of Rashba and Dresselhaus terms,
while in section~\ref{sec:hamiltonian} we discuss the independent-atom
(non-interacting) Hamiltonian including both the kinetic energy
and the magnetic parts.
The eigenvalues and eigenvectors of the single-atom Hamiltonian
are discussed in section~\ref{sec:excitation-spectrum}, where
a generalized helicity basis is introduced and Fermi surfaces for
various values of crossed Zeeman fields
are presented. Particular attention is paid to parity violations
for the eigenvalues and eigenvectors.
In section~\ref{sec:spectral-function}, we show the momentum
dependence for the spectral function at fixed frequency
revealing the cases of weak and strong parity violations,
while in section~\ref{sec:momentum-distribution} we describe the
the spin-resolved momentum distributions which also reveal
parity violations for various values of crossed Zeeman fields.
In section~\ref{sec:density-of-states}, we analyze the spin-dependent
density of states for a few values of the crossed Zeeman fields
and fixed ERD spin-orbit coupling, while
in section~\ref{sec:thermodynamic-properties}, we define several
thermodynamics properties which analyzed in the following sections,
such as pressure and entropy in section~\ref{sec:pressure-entropy},
chemical potential in section~\ref{sec:chemical-potential},
isothermal compressibility in section~\ref{sec:isothermal-compressibility}
and spin-polarization~\ref{sec:induced-polarization}.
Furthermore, we investigate the effects of parity violation
on the effective interactions in section~\ref{sec:effects-of-interactions}
and on the pairing temperature in section~\ref{sec:pairing-temperature}.
Lastly, we state our conclusions in section~\ref{sec:conclusions}
emphasizing that several measurements can be made to detect
parity violation in both non-interacting and interacting Fermi gases
in the simultaneous presence of an ERD spin-orbit field,
and crossed Zeeman fields $h_y$, $h_z$, where $h_y$ is related to the
detuning $\delta$ and $h_z$ to Raman coupling $\Omega$.

\section{Magnetic Hamiltonian}
\label{sec:magnetic-hamiltonian}

To describe quantum degenerate Fermi systems in the presence of artificial
spin-orbit and crossed artificial Zeeman fields, we discuss first
the type of magnetic Hamiltonian to be used. Generally speaking the
coupling between magnetic fields and two spin-states is given by
$
H_{\rm mag}({\bf k}) = - h({\bf k}) \cdot \bm{\sigma},
$
where ${\bf h}({\bf k})$ is the effective magnetic field (including Zeeman
and spin-orbit) and $\bm{\sigma} = (\sigma_x, \sigma_y, \sigma_z)$, with
$\sigma_i$ being the Pauli matrices. This leads to the magnetic Hamiltonian
\begin{equation}
H_{\rm mag} ({\bf k})
=
\begin{pmatrix}
h_z({\bf k}) & h_x({\bf k}) - i h_y({\bf k}) \\
h_x({\bf k}) + i h_y({\bf k}) & -h_z({\bf k})
\end{pmatrix}.
\end{equation}

Although there are many possible forms of fictitious magnetic fields that
can be created in the laboratory. We discuss here a few simpler magnetic
field configurations. In principle, the external artificial Zeeman field
${\bf h}_{\rm ZE}$ can have three components $( h_x, h_y, h_z )$,
in practice, we set $h_x = 0$, and $(h_y, h_z) \ne 0$ because in
current experimental setups $h_y = \delta/2$, where $\delta$ is the laser
detuning, and $h_z = \Omega/2$, where $\Omega$ is the strength of Raman
coupling field.

In addition, there are many possible types of spin-orbit contributions
where ${\bf h}_{SO}$ can have three components
$\left( h_{SO,x}({\bf k}), h_{SO,y} ({\bf k}), h_{SO,z} ({\bf k}) \right)$.
However, we consider particular forms of spin-orbit fields, which can be more
easily realized in practice. The first type has the Dresselhaus form
$
{\bf h}_{D} ({\bf k})
=
v_{D}
\begin{pmatrix}
k_y, & k_x, & 0
\end{pmatrix},
$
where $v_D$ measures the strength of the Dresselhaus field in units
of velocity. The corresponding Hamiltonian for such field is
\begin{equation}
\label{eqn:dresselhaus-only}
{\bf H}_{D}({\bf k})
=
-v_D
\begin{pmatrix}
0 & k_y - i k_x \\ k_y +ik_x & 0
\end{pmatrix}.
\end{equation}
The second type has the Rashba form
$
{\bf h}_R ({\bf k})
=
v_R
\begin{pmatrix}
-k_y, & k_x, & 0
\end{pmatrix},
$
where $v_R$ measures the strength of the Rashba field in units of velocity.
The corresponding Rashba Hamiltonian is
\begin{equation}
\label{eqn:rashba-only}
{\bf H}_R({\bf k})
=
v_R
\begin{pmatrix}
0 & k_y + i k_x \\ k_y -ik_x & 0
\end{pmatrix}.
\end{equation}
Either the Dresselhaus or the Rashba forms require spin-orbit fields
along the $x$ and $y$ directions, which to be produced experimentally
demand two orthogonal Raman setups, such that momentum transfer
occurs in two perpendicular directions. A linear
combination of two forms leads to the field
$
{\bf h}_{RD} ({\bf k})
=
\begin{pmatrix}
v_{RD_-} k_y, &
v_{RD_+} k_x, &
0
\end{pmatrix},
$
where the velocities
$
v_{RD_\pm} =
( v_D \pm v_R ).
$
The corresponding Hamiltonian for such linear combination is
\begin{equation}
\label{eqn:rashba-dresselhaus}
{\bf H}_{RD}({\bf k})
=
\begin{pmatrix}
0 &
v_{RD_-} k_y + i v_{RD_+}  k_x \\
v_{RD_-}  k_y - i v_{RD_+} k_x
& 0
\end{pmatrix}.
\end{equation}

The simplest type of spin-orbit field that can been created in
the laboratory has the equal-Rashba-Dresselhaus (ERD) form
$
{\bf h}_{ERD} ({\bf k})
=
v (0, k_x, 0),
$
where $v_{RD_-} = 0$, and $v_{RD_+} = v$, or equivalently
$v_D = v_R = v/2$. The
corresponding Hamiltonian for the ERD spin-orbit field has the simple form
\begin{equation}
\label{eqn:equal-rashba-dresselhaus}
{\bf H}_{ERD}({\bf k})
=
v
\begin{pmatrix}
0 &  i k_x \\ -ik_x & 0
\end{pmatrix}.
\end{equation}

Taking into account an arbitrary superposition of Rashba and Dresselhaus
spin-orbit coupling and a general uniform Zeeman field
${\bf h}_{ZE} = \left( h_x, h_y, h_z \right)$, we can write the
Zeeman-spin-orbit Hamiltonian as
\begin{equation}
{\bf H}_{ZSO}({\bf k})
=
-
\begin{pmatrix}
h_{\parallel} & h_\perp({\bf k}) \\
h_\perp^*({\bf k}) & -h_{\parallel}
\end{pmatrix},
\end{equation}
where the parallel component of the total field is $h_{\parallel} = h_z$
and the transverse component is
$$
h_\perp({\bf k})
=
\left[
h_x + h_{RD,x} ({\bf k}) - i (h_y + h_{RD,y}({\bf k})
\right],
$$
where
$
h_{RD,x} ({\bf k}) = v_{RD_-} k_y
$
and
$
h_{RD,y} ({\bf k}) = v_{RD_+} k_x.
$

Having presented the magnetic Hamiltonian, we discuss next the Hamiltonian
including the kinetic energy of the atoms.

\section{Hamiltonian}
\label{sec:hamiltonian}

The Hamiltonian for non-interacting ultra-cold fermions with identical
masses $m$ in the presence of spin-orbit and crossed Zeeman fields
can be written in second quantization as
\begin{equation}
\mathcal{H}
=
\sum_{\bf k} \Psi^\dagger({\bf k}) {\bf H}({\bf k}) \Psi({\bf k}),
\end{equation}
where the spinor
$
\Psi^\dagger({\bf k})
=
\left(
c_{{\bf k}\uparrow}^\dagger,
c_{{\bf k}\downarrow}^\dagger
\right)
$
describes the creation of fermion states with momentum ${\bf k}$ and
spin $\uparrow$ or $\downarrow$. Such Hamiltonian describes two hyperfine
states of Fermi atoms such as $^6{\rm Li}$ or $^{40}{\rm K}$,
and the corresponding Hamiltonian matrix is
\begin{equation}
\label{eqn:hamiltonian-matrix-non-interacting}
{\bf H} ({\bf k}) =
\begin{pmatrix}
{\widetilde K}_\uparrow ({\bf k}) & -h_\perp({\bf k})\\
-h_\perp^*({\bf k}) & {\widetilde K}_\downarrow ({\bf k})
\end{pmatrix},
\end{equation}
where
$
{\widetilde K}_\sigma ({\bf k}) = k^2/(2m) - \mu_{\sigma}
$
represents the kinetic energy of a fermion with mass $m$,
momentum ${\bf k}$ and spin $\sigma$ with respect to the
chemical potential $\mu_{\sigma}$.

We define the variables
$
{\widetilde K}_{\pm}
=
\left(
{\widetilde K}_\uparrow \pm {\widetilde K}_\downarrow
\right)
/
2
$
and the chemical potentials
$
\mu_{\pm}
=
\left(
\mu_{\uparrow} \pm \mu_{\downarrow}
\right)
/
2,
$
and notice that
$
{\widetilde K}_+
=
\vert {\bf k} \vert^2/(2m)
-
\mu_+
=
\xi_{{\bf k} +}
$
plays the role of the average kinetic energy,
while
$
{\widetilde K}_-
=
-(\mu_- + h_z)
\equiv
-h_{\parallel},
$
plays the role of the parallel Zeeman field $h_{\parallel}$
including the external field $h_z$ and the internal field
$\mu_-$ due to a possible initial population imbalance.
In the limit that there is zero initial population imbalance,
the chemical potentials can be set to $\mu_+ \to \mu$ and $\mu_- \to 0$, while
the kinetic energies reduce to ${\widetilde K}_+ ({\bf k}) \to \xi_{\bf k}$ and
${\widetilde K}_- ({\bf k}) \to -h_\parallel = - h_z$,
where $\xi_{\bf k} = \vert {\bf k} \vert^2/(2m) - \mu$.

The non-interacting Hamiltonian has the simpler form
$
{\bf H}_0 ({\bf k})
=
\xi_{\bf k} {\bf 1} - h_\parallel  \sigma_z
- h_x({\bf k})  \sigma_x - h_y ({\bf k}) \sigma_y
$
which can be re-expressed as
\begin{equation}
\label{eqn:hamiltonian-non-interacting}
{\bf H}_0 ( {\bf k})
=
\begin{pmatrix}
\xi_{\bf k} - h_{\parallel} & -h_\perp({\bf k}) \\
-h_\perp^*({\bf k}) & \xi_{\bf k} + h_{\parallel}
\end{pmatrix},
\end{equation}
where
$
h_\perp ({\bf k})
=
h_x ({\bf k})
-
i h_y ({\bf k}).
$
In the remainder of the manuscript, we will discuss this
explicit form of the Hamiltonian and some interesting
consequences.

We define the total number of fermions as
$N = N_\uparrow + N_\downarrow$,
and choose our energy, velocity and momentum
scales through the Fermi momentum $k_{F}$ defined from the
total density of fermions
$
n
=
n_\uparrow
+
n_\downarrow
=
k_{F}^3/(3\pi^2),
$
where $n = N/V$ and $n_s = N_s/V$ with $s = (\uparrow, \downarrow)$.
This choice leads to the Fermi energy
$\epsilon_{F} = k_{F}^2/2m$
and to the Fermi velocity $v_{F} = k_{F}/m$,
which are the energy and velocity scales used throughout the
manuscript.

\section{Excitation Spectrum}
\label{sec:excitation-spectrum}

Now, let us introduce the unitary matrix
${\bf U}_{\bf k}$ that diagonalizes the
Hamiltonian ${\bf H_0} ({\bf k})$,
such that
\begin{equation}
{\bf E} ({\bf k})
=
{\bf U}^\dagger_{\bf k}
{\bf H}_0 ({\bf k})
{\bf U}_{\bf k}
\end{equation}
is a diagonal matrix containing the eigenvalues of ${\bf H}_0 ({\bf k})$.
The corresponding eigenvectors are the spinors
$
\Phi ({\bf k})
=
{\bf U}_{\bf k}^{\dagger}
\Psi ({\bf k}),
$
where the unitary matrix ${\bf U}_{\bf k}$ has
a momentum-dependent SU(2) form and can
be written as
\begin{equation}
\label{unitary-matrix}
{\bf U}_{\bf k}
=
\begin{pmatrix}
u_{\bf k} & v_{\bf k} \\
-v^*_{\bf k} & u_{\bf k}
\end{pmatrix},
\end{equation}
where the normalization condition
$
|u_{\bf k}|^2
+
|v_{\bf k}|^2
=
1
$
is imposed to satisfy the unitarity condition
$
{\bf U}_{\bf k}^\dagger
{\bf U}_{\bf k}
=
{\bf 1}
$
leading to the following expressions
\begin{equation}
u_{\bf k}
=
\sqrt{
\frac{1}{2}
\left(
1
+
\frac{h_{\parallel}}
{\lvert
{\bf h}_{{\rm eff}}({\bf k})
\rvert}
\right)
},
\end{equation}
where $u_{\bf k}$ is taken to be real without
loss of generality and
\begin{equation}
v_{\bf k}
=
- e^{i\varphi_{\bf k}}
\sqrt{
\frac{1}{2}
\left(
1 -
\frac{h_{\parallel}}
{\lvert
{\bf h}_{\rm {eff}}({\bf k})
\rvert}
\right)
},
\end{equation}
is a complex function where the phase $\varphi_{\bf k}$
is defined via
$
h_\perp({\bf k})
=
\vert
h_\perp({\bf k})
\vert
e^{i\varphi_{\bf k}},
$
leading to
$
\varphi_{\bf k}
=
{\rm Arg}
\left[
h_\perp ({\bf k})
\right]
$
.

The eigenvalues of ${\bf H}_0 ({\bf k})$ emerge as
\begin{equation}
{\bf E} ({\bf k})
=
\begin{pmatrix}
\xi_{\Uparrow} ({\bf k}) & 0 \\
0  & \xi_{\Downarrow} ({\bf k})
\end{pmatrix},
\end{equation}
where
$
\xi_\Uparrow ({\bf k})
=
\xi_k
-
\lvert
{\bf h}_{\rm eff} ({\bf k})
\rvert
$
is the eigenvalue where the momentum dependent
effective field ${\bf h}_{\rm eff} ({\bf k})$ is aligned with
the spin $\Uparrow$, and
$
\xi_\Downarrow ({\bf k})
=
\xi_{\bf k}
+
\lvert
{\bf h}_{\rm eff} ({\bf k})
\rvert
$
is the eigenvalue where the momentum dependent
effective field ${\bf h}_{\rm eff} ({\bf k})$
is aligned with the spin $\Downarrow$.
Here,
$
\lvert
{\bf h}_{\rm eff} ({\bf k})
\rvert
=
\sqrt{
h_{\parallel}^2
+
\lvert
h_\perp({\bf k})
\rvert^2
}
$
is the magnitude of the effective field.
The respective eigenvectors are
\begin{equation}
\Phi_{\Uparrow}({\bf k})
=
u_{{\bf k}}  c_{{\bf k} \uparrow}
-
v_{{\bf k}}  c_{{\bf k} \downarrow}
\end{equation}
corresponding to the state $\vert {\bf k}, \Uparrow \rangle$,
and
\begin{equation}
\Phi_{\Downarrow}({\bf k})
=
v_{{\bf k}}^*  c_{{\bf k} \uparrow}
+
u_{{\bf k}}  c_{{\bf k} \downarrow}
\end{equation}
corresponding to the state $\vert {\bf k}, \Downarrow \rangle$.

\begin{figure} [htb]
\includegraphics[width = 1.0\linewidth]{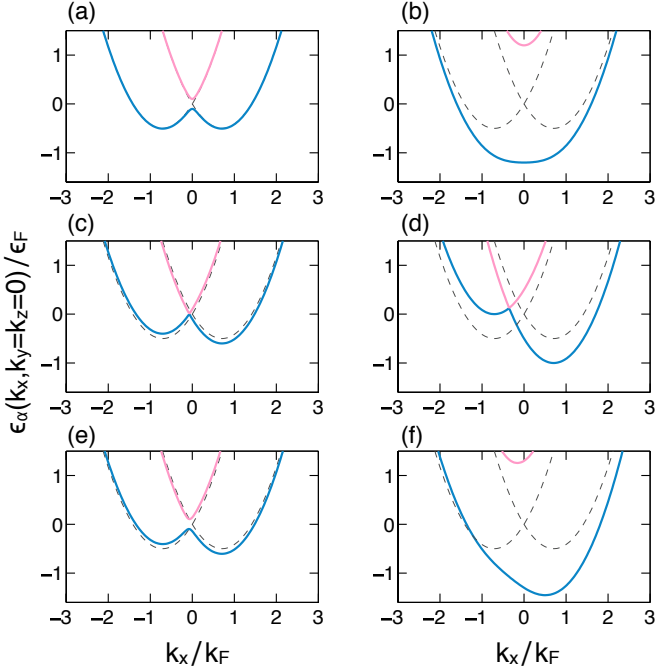}
\caption{
\label{fig:one}
(color online)
Helicity energy dispersions
$\epsilon_\Uparrow ({\bf k})/\epsilon_F$
(blue line) and $\epsilon_\Downarrow ({\bf k})/\epsilon_F$ (magenta line)
versus momentum $k_x/k_F$ with $k_y = k_z = 0$ for ERD spin-orbit coupling
$v/v_F = 0.71$.
For reference, the black dashed lines show the helicity bands
for $v/v_F = 0.71$ with $h_z/\epsilon_F = h_y/\epsilon_F = 0$.
The Zeeman fields are
(a) $h_y/\epsilon_F = 0$ and $h_z/\epsilon_F = 0.1$,
(b) $h_y/\epsilon_F = 0$ and $h_z/\epsilon_F = 1.2$,
(c) $h_y/\epsilon_F = 0.1$ and $h_z/\epsilon_F = 0$,
(d) $h_y/\epsilon_F = 0.5$ and $h_z/\epsilon_F = 0$,
(e) $h_y/\epsilon_F = h_z/\epsilon_F = 0.1$,
(f) $h_y/\epsilon_F = 0.5$ and $h_z/\epsilon_F = 1.2$.
}
\end{figure}

For initially population balanced systems in the ERD case,
with the Zeeman field having only $h_y$ and $h_z$ components,
the magnitude of the effective magnetic field
$
\vert h_{\rm eff} ({\bf k}) \vert
=
\sqrt{
h_z^2
+
\left(h_y + v k_x\right)^2}
$
does not have well defined parity, which implies that
the same is true for the eigenvalues $\xi_{\alpha} ({\bf k})$.
If either $h_y = 0$ (zero detuning) or $v = 0$ (zero spin-orbit
coupling), parity is restored for the eigenvalues.
This can be seen in Fig.~\ref{fig:one},
where the energy dispersions
$
\epsilon_\Uparrow ({\bf k})
=
\epsilon_{\bf k}
-
\vert h_{\rm eff} ({\bf k}) \vert
$
and
$
\epsilon_\Downarrow ({\bf k})
=
\epsilon_{\bf k}
+
\vert h_{\rm eff} ({\bf k}) \vert
$
are shown for several cases.
In all panels of Fig.~\ref{fig:one} the
dashed lines represent the case where
the ERD spin orbit is finite $(v/v_F = 0.71)$,
but the Zeeman fields are zero $(h_z = h_y = 0)$.
This case corresponds to shifted parabolic bands
$
\epsilon_\Uparrow ({\bf k}) = k_x^2/(2m) - \vert v k_x \vert +
k_\rho^2/(2m)
$
and
$
\epsilon_\Downarrow ({\bf k}) = k_x^2/(2m) + \vert v k_x \vert + k_\rho^2/(2m),
$
where $k_\rho^2 = k_y^2 + k_z^2$. The lower helicity band
$\epsilon_\Uparrow ({\bf k})$ has two minima, which
occur at $(k_x = m \vert v \vert, k_y = 0, k_z = 0)$ and
$(k_x = - m\vert v \vert, k_y = 0, k_z = 0)$, respectively,
while the upper helicity band $\epsilon_\Uparrow ({\bf k})$
has only one minimum occurring at $(k_x = 0, k_y = 0, k_z = 0)$.
This can be seen by completing the squares and rewriting
the dispersions of the helicity bands as
$
\epsilon_\Uparrow ({\bf k})
=
(\vert k_x \vert - m \vert v \vert)^2/(2m) - \epsilon_v + k_\rho^2/(2m)
$
and
$
\epsilon_\Downarrow ({\bf k})
=
(\vert k_x \vert + m \vert v \vert)^2/(2m) - \epsilon_v + k_\rho^2/(2m),
$
where $\epsilon_v = mv^2/2$ is the characteristic kinetic energy
associated with the spin-orbit coupling strength $v$, which
has units of velocity.

In Fig.~\ref{fig:one}a, we show the case of
$v/v_F = 0.71$, for $h_y = 0$ and $h_z/\epsilon_F = 0.1$,
and in Fig.~\ref{fig:one}b, we show the case of
$v/v_F = 0.71$, for $h_y = 0$ and $h_z/\epsilon_F = 1.2$.
In both of these cases the energy dispersions are parity
preserving, and the main fundamental difference
between Fig.~\ref{fig:one}a and Fig.~\ref{fig:one}b is that the lower helicity
band $\epsilon_{\Uparrow} ({\bf k})$
has two minima at finite
$k_x = \pm \vert v \vert^{-1}\sqrt{4\epsilon_v^2 - h_z^2}$ as seen in
Fig.~\ref{fig:one}a, but a single minimum at $k_x = 0$
as seen in Fig.~\ref{fig:one}b. While the
upper helicity band $\epsilon_{\Downarrow} ({\bf k})$
has only a single minimum at $k_x = 0$ in both Fig.~\ref{fig:one}a and
Fig.~\ref{fig:one}b.
The two minima in the lower helicity band for Fig.~\ref{fig:one}a occur
only for low Zeeman fields $\vert h_z \vert < 2\epsilon_v$.
In Figs.~\ref{fig:one}c and~\ref{fig:one}d,
we show the helicity bands for $h_z = 0$, but $h_y/\epsilon_F = 0.1$
and $h_y/\epsilon_F = 0.5$, respectively.  In these cases, the energy
dispersions are
$
\epsilon_\Uparrow ({\bf k})
=
k_x^2/(2m)  - \vert h_y + v k_x \vert + k_\rho^2/(2m)
$
and
$
\epsilon_\Downarrow ({\bf k})
=
k_x^2/(2m) + \vert h_y + v k_x \vert + k_\rho^2/(2m)
$
and do not even parity as it is standard,
since
$
\epsilon_\alpha ({\bf k}) \ne \epsilon_\alpha (-{\bf k}).
$
These dispersions are compared to the dispersions for
the case of $h_y/\epsilon_F = h_z/\epsilon_F = 0$, but finite
$v/v_F$, which are shown as dashed black lines in Fig.~\ref{fig:one}.
The last examples described in Fig.~\ref{fig:one}e and
Fig.~\ref{fig:one}f
correspond to cases where both $h_y$ and $h_z$ are non-zero, having
values $h_y/\epsilon_F = h_z/\epsilon_F = 0.1$ for (e)
and $h_y/\epsilon_F = 0.5$, $h_z/\epsilon_F = 1.2$ for (f). Notice the
presence of two minima for the lower helicity band in (e),
and the existence of only one minimum for the lower helicity band in (f),
but in both cases parity (inversion symmetry) is violated.

\begin{figure} [htb]
\includegraphics[width = 1.0\linewidth]{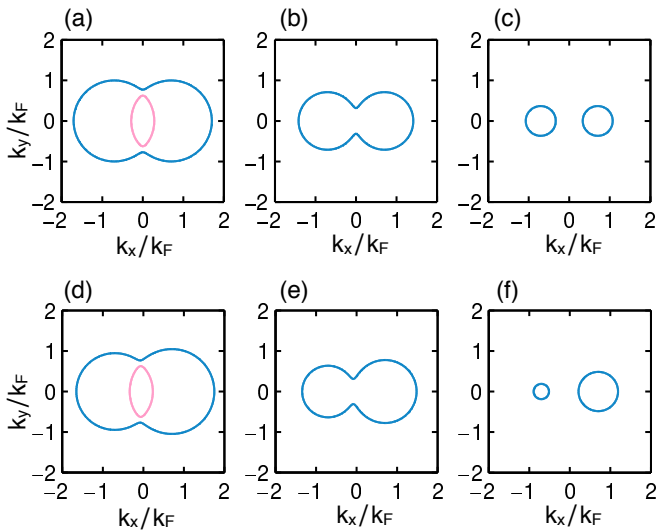}
\caption{
\label{fig:two}
(color online)
Cross sections of Fermi surfaces in the $k_x$-$k_y$ plane
at $k_z = 0$ for ERD spin-orbit coupling $v/v_F = 0.71$.
In (a) through (c), the Zeeman fields are
$h_y/\epsilon_F = 0$ and $h_z/\epsilon_F = 0.1$, where
parity symmetry is preserved, and the different values
of the chemical potential and induced polarization are
(a) $\mu/\epsilon_F = 0.495$,  $P_{\rm ind} = 0.121$,
(b) $\mu/\epsilon_F = 0$,  $P_{\rm ind} = 0.147$,
(c) $\mu/\epsilon_F = -0.370$, $P_{\rm ind} = 0.106$.
In (d) through (f) the Zeeman fields are
$h_y/\epsilon_F =  h_z/\epsilon_F = 0.1$, where
parity symmetry is not preserved, and the different values
of the chemical potential and induced polarization are
(d) $\mu/\epsilon_F = 0.492$, $P_{\rm ind} = 0.121$,
(e) $\mu/\epsilon_F = 0$, $P_{\rm ind} = 0.143$,
(f) $\mu/\epsilon_F = -0.370$, $P_{\rm ind} = 0.100$.
The blue and magenta lines indicate the energy contours
of the lower $\epsilon_{\Uparrow} ({\bf k})$ and upper
$\epsilon_{\Uparrow} ({\bf k})$ helicity bands, respectively.
}
\end{figure}

In Fig.~\ref{fig:two}, we show cross sections of the Fermi surfaces (FS)
in the $k_x$-$k_y$ plane at $k_z = 0$ for ERD spin-orbit coupling
and a few values of crossed Zeeman fields. The Fermi surfaces
have rotational symmetry about the $k_x$ axis, that is, in the
$k_y$-$k_z$ plane, and their full three-dimensional structure
can be visualized using this property.
We show in Figs.~\ref{fig:two}a-c the case for Zeeman fields
$h_y/\epsilon_F = 0$ and $h_z/\epsilon_F = 0.1$,
where the Fermi surfaces exhibit parity (or inversion) symmetry.
These parameters correspond to the helicity bands shown
in Fig.~\ref{fig:one}a, where the lower helicity band has two minima.
While we show in Figs.~\ref{fig:two}d-f the case
for Zeeman fields $h_y/\epsilon_F = h_z/\epsilon_F = 0.1$,
where the corresponding Fermi surfaces do not have well defined parity
or inversion symmetry. The specific values of the chemical potential
$\mu$ and induced polarization
$
P_{\rm ind}
=
( N_\uparrow - N_\downarrow )
/
( N_\uparrow + N_\downarrow )
$
are indicated in the captions.

In Fig.~\ref{fig:two}, notice also that as the chemical potential
$\mu$ is changed below the bottom of the helicity band
$\epsilon_\Downarrow ({\bf k})$, the central pocket of the Fermi surface
disappears (see magenta surface near zero momentum in Fig.~\ref{fig:two}).
Further lowering of the chemical potential leads to the crossing
of a local maximum of the helicity band $\epsilon_\Uparrow ({\bf k})$,
where the residual Fermi surface break into two pockets (see blue surfaces
in Fig.~\ref{fig:two}).
This is reminiscent of the Lifshitz transition~\cite{lifshitz-1960}
in non-interacting metals, also called metal-to-metal or
conductor-to-conductor transition, where under pressure or another external
parameter the Fermi surface of the system changes topology producing a major
rearrangement of momentum states which lead to a drastic change in the
density of states of the system. The thermodynamic potential $\Omega$ in the
vicinity of the usual Lifshitz transition behaves as
$
\Omega
=
\Omega_{\rm reg}
+
\alpha \vert \mu - \mu_c \vert^{5/2},
$
where $\Omega_{\rm reg}$ is the
regular (analytic) part, and $\alpha$ is the prefactor of the non-analytic
component. The isothermal compressibility is related to the second-derivative
of the thermodynamic potential with respect to the chemical potential and
behaves as
$
\kappa_T
=
\kappa_{T, {\rm reg}}
+
\beta \vert \mu - \mu_c \vert^{1/2},
$
where $\kappa_{T, {\rm reg}}$ is the regular part and
$\beta$ is the coefficient of the non-analytic component.
According to Ehrenfest's classification of phase transitions,
the non-analyticity manifests itself only in the third derivative
and is a third-order phase transition. However, the Lifshitz transition
is more commonly called the $2$-$1/2$ order transition in
allusion to the specific $5/2$ power-law non-analyticity of $\Omega$
in three dimensions. This topological transition is not characterized
under Landau's symmetry-based classification, since no symmetry is broken
in the Lifshitz case. This {\it trivial} Lifshitz transition can be seen in
Fig.~\ref{fig:two} both for the parity-preserving and parity-violating
examples.

\begin{figure} [htp]
\includegraphics[width = 1.0\linewidth]{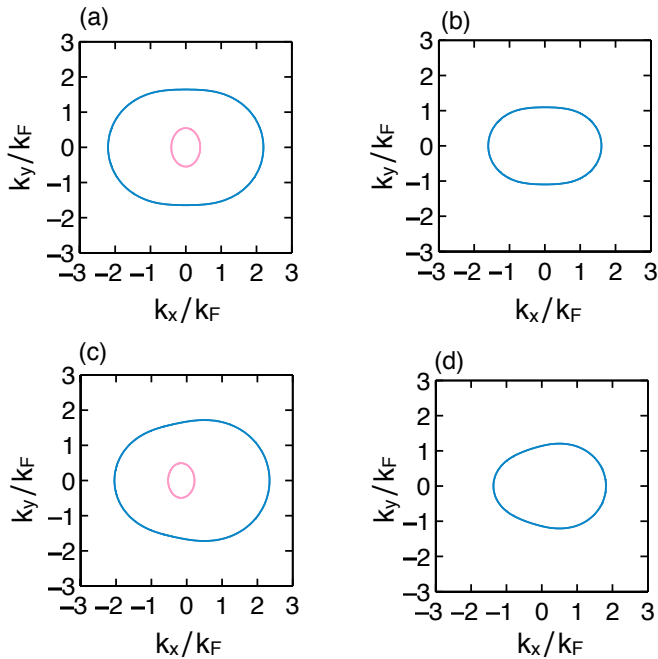}
\caption{
\label{fig:three}
(color online)
Cross sections of Fermi surfaces in the $k_x$-$k_y$ plane
at $k_z = 0$ for ERD spin-orbit coupling $v/v_F = 0.71$.
In (a) and (b), the Zeeman fields are
$h_y/\epsilon_F = 0$ and $h_z/\epsilon_F = 1.2$,
with parity symmetry being preserved. The values
of the chemical potential and induced polarization are
(a) $\mu/\epsilon_F = 1.5$, $P_{\rm ind} = 0.692$;
(b) $\mu/\epsilon_F = 0$, $P_{\rm ind} = 0.801$.
In (c) and (d), the Zeeman fields are
$h_y/\epsilon_F = 0.5$, $h_z/\epsilon_F = 1.2$ and
parity symmetry is not preserved. The values of the chemical potential
and induced polarization are
(c) $\mu/\epsilon_F = 1.5$, $P_{\rm ind} = 0.673$;
(d) $\mu/\epsilon_F = 0$, $P_{\rm ind} = 0.739$.
The blue and magenta lines indicate the energy contours
of the lower $\epsilon_{\Uparrow} ({\bf k})$ and upper
$\epsilon_{\Uparrow} ({\bf k})$ helicity bands, respectively.
}
\end{figure}

In Fig.~\ref{fig:three},
we also show cross sections of the Fermi surfaces (FS)
in the $k_x$-$k_y$ plane at $k_z = 0$ for ERD spin-orbit coupling
$v/v_F = 0.71$ and a few values of crossed Zeeman fields. The Fermi surfaces
have also rotational symmetry about the $k_x$ axis, that is, in the
$k_y$-$k_z$ plane.
We show in Figs.~\ref{fig:three}a-b the case for Zeeman fields
$h_y/\epsilon_F = 0$ and $h_z/\epsilon_F = 1.2$,
where the Fermi surfaces exhibit parity (or inversion) symmetry.
These parameters correspond to the helicity bands shown
in Fig.~\ref{fig:one}b, where the lower helicity band has only one minimum.
While we show in Figs.~\ref{fig:three}c-d the cases of
the Zeeman fields $h_y/\epsilon_F = 0.5$ and $h_z/\epsilon_F = 1.2$
where the corresponding Fermi surfaces do not have well defined parity
or inversion symmetry. The values of the
chemical potential $\mu$ and induced polarization $P_{\rm ind}$ are
$\mu/\epsilon_F = 1.5$, $P_{\rm ind} = 0.692$ for Fig.~\ref{fig:three}a;
$\mu/\epsilon_F = 0$, $P_{\rm ind} = 0.801$ for Fig.~\ref{fig:three}b;
$\mu/\epsilon_F = 1.5$, $P_{\rm ind} = 0.673$ for Fig.~\ref{fig:three}c;
and
$\mu/\epsilon_F = 0$, $P_{\rm ind} = 0.739$ for Fig.~\ref{fig:three}d.
Notice that there is a fundamental difference between the Fermi surfaces in
Figs.~\ref{fig:two} and~\ref{fig:three} in connection
with their topology. The lower helicity band $\epsilon_\Uparrow ({\bf k})$
can have two simply connected FS for the parameters of Fig.~\ref{fig:two},
but only one simply connected FS for the parameters of Fig.~\ref{fig:three}.
This means that there is only one {\it trivial} Lifshitz transition
in the case of Fig.~\ref{fig:three}, while there are two {\it trivial}
Lifshitz transitions in the case of Fig.~\ref{fig:two}.
We call this transition for non-interacting systems {\it trivial}
to contrast it with a more exotic, but related  topological transition
that can occur in $p$-wave~\cite{volovik-1992, botelho-2005a}
or $d$-wave~\cite{duncan-2000, botelho-2005b} superfluids, where
interactions play a fundamental role.

A further characterization of the parity violation present in fermion
systems with spin-orbit coupling and crossed Zeeman fields can be made by
analyzing additional spectroscopic quantities such as the spectral function
to be discussed next.

\section{Spectral Function}
\label{sec:spectral-function}

A very useful tool to probe momentum-resolved properties is the use
of radio-frequency (RF) spectroscopy that can extract
the spectral function~\cite{jin-2010} yielding similar measurements
to those encountered in photoemission spectroscopy of
condensed matter physics. The resolvent (or Green) operator matrix
is defined as
\begin{equation}
{\bf G} ({\bf k}, z)
=
\frac{1}{ z{\bf 1} - {\bf H} ({\bf k}) }
\end{equation}
in momentum-frequency space.
In the present case, the diagonal components of
${\bf G}({\bf k}, z)$ are
\begin{equation}
G_{\uparrow}({\bf k},z)
=
\frac{u_{\bf k}^2}{z-\xi_\Uparrow({\bf k})}
+
\frac{\lvert v_{\bf k} \rvert^2}{ z -\xi_\Downarrow({\bf k}) }
\end{equation}
for the spin $\uparrow$ component and
\begin{equation}
G_{\downarrow}({\bf k},z)
=
\frac{\lvert v_{\bf k} \rvert^2}{z -\xi_\Uparrow ({\bf k})}
+
\frac{u_{\bf k}^2}{z -\xi_\Downarrow ({\bf k})}
\end{equation}
for the spin $\downarrow$ component.
The corresponding spectral function is
$
A_s ({\bf k}, \omega)
=
- \pi^{-1}
{\rm {Im}}
\left[
G_s ({\bf k}, \omega + i\delta)
\right],
$
which in terms of the coherence factor $u_{\bf k}$ and
$v_{\bf k}$ becomes
\begin{equation}
A_\uparrow ({\bf k},\omega)
=
u_{\bf k}^2  \delta (\omega - \xi_\Uparrow ({\bf k}) )
+
|v_{\bf k}|^2  \delta (\omega - \xi_\Downarrow ({\bf k}) )
\end{equation}
for the up-spin $({\uparrow})$ component, and
\begin{equation}
A_\downarrow ({\bf k}, \omega)
=
\vert
v_{\bf k}
\vert^2
\delta(\omega - \xi_\Uparrow ({\bf k}) )
+
u_{\bf k}^2  \delta (\omega - \xi_\Downarrow ({\bf k}) )
\end{equation}
for the down-spin $(\downarrow)$ component.

\begin{figure} [htb]
\includegraphics[width = 1.0\linewidth]{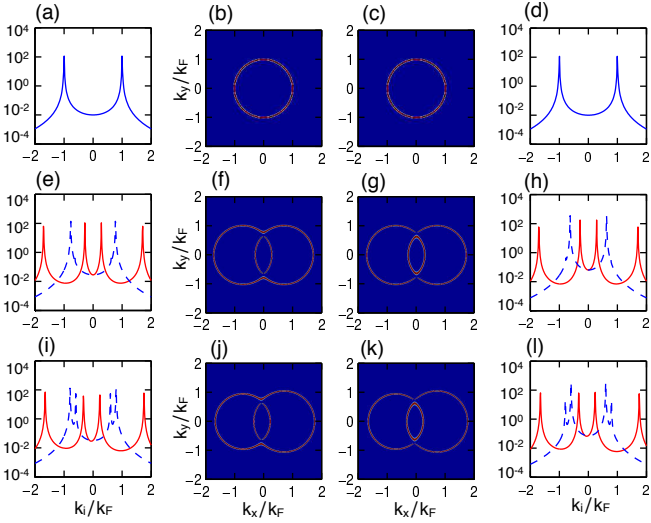}
\caption{
\label{fig:four}
(color online)
Finite temperature $(T/\epsilon_F = 0.05)$
dimensionless spectral functions $2\pi \epsilon_F A_s({\bf k}, \omega)$
at $\omega = \mu$ for parameters
(a)-(d) $v/v_F = 0$, $h_y/\epsilon_F = h_z/\epsilon_F = 0$
and $\mu/\epsilon_F =1.0$;
(e)-(h) $v/v_F = 0.71$, $h_y/\epsilon_F = 0$, $h_z/\epsilon_F = 0.1$,
and $\mu/\epsilon_F = 0.495$;
(i)-(l) $v/v_F = 0.71$, $h_y/\epsilon_F = h_z/\epsilon_F = 0.1$
and $\mu/\epsilon_F = 0.492$.
The blue dashed lines are cuts along the direction
$(k_x = 0, k_y, k_z = 0)$ corresponding
to $2\pi \epsilon_F A_s (k_x = 0, k_y, k_z = 0, \omega = \mu)$, and
the red solid lines
are cuts along the direction $(k_x, k_y = 0, k_z = 0)$
corresponding to $2\pi \epsilon_F A_s (k_x, k_y = 0, k_z = 0, \omega = \mu)$.
The two left-most columns describe
the $A_\uparrow ({\bf k})$ component, and the two right-most columns
correspond to the $A_\downarrow ({\bf k})$ component.
}
\end{figure}

The spectral functions $A_s({\bf k}, \omega)$
at frequency $\omega = \mu$ and temperature $T/\epsilon_F = 0.05$
are shown in Fig.~\ref{fig:four} for some values of the
crossed Zeeman fields and spin-orbit coupling.
A small energy broadening $\delta/\epsilon_F = 0.01$ is included
and a logarithmic scale is used to help visualization.
In the relevant panels
of Fig.~\ref{fig:four}, the blue dashed lines
represent $A_s (k_x = 0, k_y, k_z = 0, \omega = \mu)$,
and the red solid lines represent $A_s (k_x, k_y = 0, k_z = 0, \omega = \mu)$.
Additionally, the two left-most columns correspond to
the $A_\uparrow ({\bf k})$ component, and the two right-most columns describe
the $A_\downarrow ({\bf k})$ component.
In Figs.~\ref{fig:four}(a)-(d) the Zeeman fields are
$h_y/\epsilon_F = h_z/\epsilon_F = 0$, the ERD coupling is $v/v_F = 0$,
the chemical potential is
$\mu/\epsilon_F = 1.0$, and the induced polarization $P_{\rm ind} = 0$.
In Figs.~\ref{fig:four}(e)-(h) the Zeeman fields are
$h_y/\epsilon_F = 0$, $h_z/\epsilon_F = 0.1$, the ERD coupling is
$v/v_F = 0.71$, and the chemical potential is
$\mu/\epsilon_F = 0.495$, and the induced polarization is $P_{\rm ind} = 0.121$.
In these two cases parity is preserved, since
the eigenvalues $\epsilon_\alpha ({\bf k})$ and
coherence factors $u_{\bf k}$ and $v_{\bf k}$ are invariant under
momentum inversion for $h_y/\epsilon_F = 0$.
In Figs.~\ref{fig:four}(i)-(l) the Zeeman fields are
$h_y/\epsilon_F = h_z/\epsilon_F = 0.1$, the ERD coupling is $v/v_F = 0.71$,
the chemical potential is $\mu/\epsilon_F = 0.492$, and the induced
polarization is $P_{\rm ind} = 0.120$.
Parity is not preserved in the last case
since $(h_y/\epsilon_F \ne 0$). This is reflected in the absence of
inversion symmetry, which is noticeable but weak.

\begin{figure} [htb]
\includegraphics[width = 1.0\linewidth]{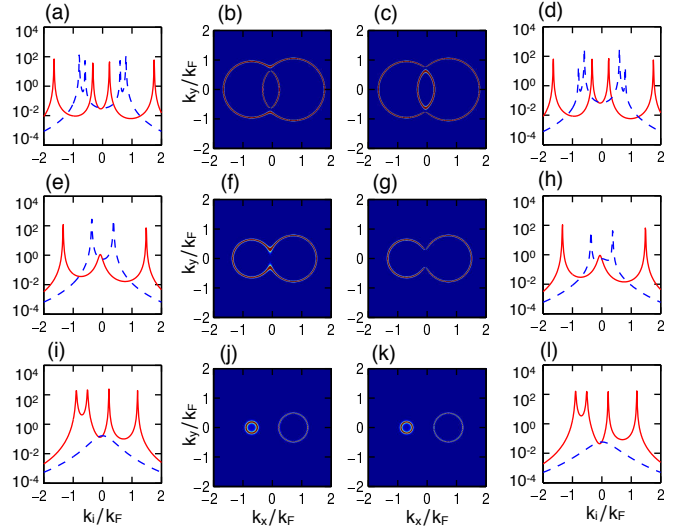}
\caption{
\label{fig:five}
(color online) Finite temperature $(T/\epsilon_F = 0.05)$
dimensionless spectral functions $2\pi \epsilon_F A_s({\bf k}, \omega)$
at $\omega = \mu$ for ERD spin-orbit coupling
$v/v_F = 0.71$ and Zeeman fields
$h_y/\epsilon_F = h_z/\epsilon_F = 0.1$
In panels
(a)-(d) $\mu/\epsilon_F = 0.492$;
(e)-(h) $\mu/\epsilon_F = 0$;
and
(i)-(l) $\mu/\epsilon_F = -0.37$.
The blue dashed lines are cuts along the direction
$(k_x = 0, k_y, k_z = 0)$ corresponding
to $2\pi \epsilon_F A_s (k_x = 0, k_y, k_z = 0, \omega = \mu)$, and
the red solid lines
are cuts along the direction $(k_x, k_y = 0, k_z = 0)$
corresponding to $2\pi \epsilon_F A_s (k_x, k_y = 0, k_z = 0, \omega = \mu)$.
The two left-most columns describe
the $A_\uparrow ({\bf k})$ component, and the two right-most columns
correspond to the $A_\downarrow ({\bf k})$ component.
}
\end{figure}

In Fig.~\ref{fig:five}, we show the spectral functions
$A_s({\bf k}, \omega)$ at frequency $\omega = \mu$
and temperature $T/\epsilon_F = 0.05$ for ERD spin-orbit
coupling $v/v_F = 0.71$. For varying chemical potentials,
we choose the particular values of the crossed Zeeman
fields to be $h_y/\epsilon_F = h_z/\epsilon_F = 0.1$
in order to emphasize the absence of inversion symmetry
(parity) when $h_y \ne 0$. As before, a small energy broadening
$\delta/\epsilon_F = 0.01$ is included and a logarithmic scale is
used to help visualization.
The same color convention is used in the relevant panels
of Fig.~\ref{fig:five}, where the blue dashed lines
represent $A_s (k_x = 0, k_y, k_z = 0, \omega = \mu)$,
and the red solid lines represent $A_s (k_x, k_y = 0, k_z = 0, \omega = \mu)$.
In addition, the two left-most columns describe
the $A_\uparrow ({\bf k})$ component, and the two right-most columns
correspond to the $A_\downarrow ({\bf k})$ component.
In Figs.~\ref{fig:five}(a)-(d) the chemical potential is
$\mu/\epsilon_F = 0.492$, the induced
polarization is $P_{\rm ind} = 0.120$,
and parity is weakly violated, but noticeable.
In Figs.~\ref{fig:five}(e)-(h) the chemical potential is
$\mu/\epsilon_F = 0$, the induced
polarization is $P_{\rm ind} = 0.143$,
and parity is more strongly violated.
In Figs.~\ref{fig:five}(i)-(l) the chemical potential is
$\mu/\epsilon_F = -0.37$, the induced
polarization is $P_{\rm ind} = 0.105$,
and parity is strongly violated.

In the last two cases parity is violated more strongly, since
the eigenvalues $\epsilon_\alpha ({\bf k})$ and
coherence factors $u_{\bf k}$ and $v_{\bf k}$ are not invariant under
momentum inversion for $h_y/\epsilon_F \ne 0$ and are more sensitive
to this violation for chemical potentials closer to the bottom of the
helicity bands.

\section{Momentum Distribution}
\label{sec:momentum-distribution}

To understand the momentum distribution for quantum degenerate
Fermi gases in the presence of spin-orbit and crossed Zeeman fields,
we look first at the expectation value
$
\langle
{\hat n}_{\alpha} ({\bf k})
\rangle
$
of the number operators
$
{\hat n}_{\alpha} ({\bf k})
=
\Phi_{\alpha}^\dagger ({\bf k}) \Phi_{\alpha} ({\bf k}),
$
which describes the momentum distribution
$n_{\alpha} ({\bf k})$
in the helicity basis.
In this case, the momentum distribution is
$n_{\alpha} ({\bf k}) = n_F (\xi_{\alpha} ({\bf k}))$,
where $n_F (x)$ is the Fermi function.

The momentum distributions for the original spin states
$s = (\uparrow, \downarrow)$ are defined as
$
n_s ({\bf k})
=
\langle
c_{{\bf k} s}^\dagger c_{{\bf k} s}
\rangle.
$
With the help of the unitary matrix ${\bf U} ({\bf k})$
defined in Eq.~(\ref{unitary-matrix}), which relates
the creation and annihilation operators in the helicity and
the standard spin basis the momentum distributions become
\begin{equation}
n_\uparrow ({\bf k})
=
u_{\bf k}^2  n_{F} (\xi_{\Uparrow} ({\bf k}))
+
|v_{\bf k}|^2  n_{F} (\xi_{\Downarrow} ({\bf k}))
\end{equation}
for the up-spin $(\uparrow)$ component, and
\begin{equation}
n_\downarrow ({\bf k})
=
|v_{\bf k}|^2  n_{F} ( \xi_{\Uparrow} ({\bf k}) )
+
u_{\bf k}^2  n_{F} ( \xi_{\Downarrow}({\bf k}) )
\end{equation}
for the down-spin $(\downarrow)$ component.

Such expressions can be also obtained from the
general relation
\begin{equation}
\label{eqn:md-sf-relation}
n_{s}({\bf k})
=
\int_{-\infty}^{\infty}
d\omega
n_F(\omega)
A_{s}({\bf k},\omega),
\end{equation}
between the momentum distribution
and the spectral function for fermions.

It is also convenient to obtain the momentum distribution sum
$
n_+({\bf k})
=
n_\uparrow ({\bf k})
+
n_\downarrow ({\bf k})
$
and the momentum distribution difference
$
n_-({\bf k})
=
n_\uparrow ({\bf k})
-
n_\downarrow ({\bf k}).
$
The first distribution can be written in terms of the
Fermi functions only
\begin{equation}
\label{eqn:momentum-distribution-sum}
n_+({\bf k})
=
n_{F} (\xi_{\Uparrow} ({\bf k}))
+
n_{F} (\xi_{\Downarrow} ({\bf k})),
\end{equation}
while the second distribution can be written in terms
of the Fermi functions and the components of the
effective Zeeman field
\begin{equation}
\label{eqn:momentum-distribution-difference}
n_-({\bf k})
=
\frac{ h_{\parallel} }{ \vert h_{{\rm eff}}({\bf k}) \vert }
\left[
 n_{F} (\xi_{\Uparrow} ({\bf k})
- n_{F} (\xi_{\Downarrow} ({\bf k})
\right].
\end{equation}

\begin{figure} [htb]
\includegraphics[width = 1.0\linewidth]{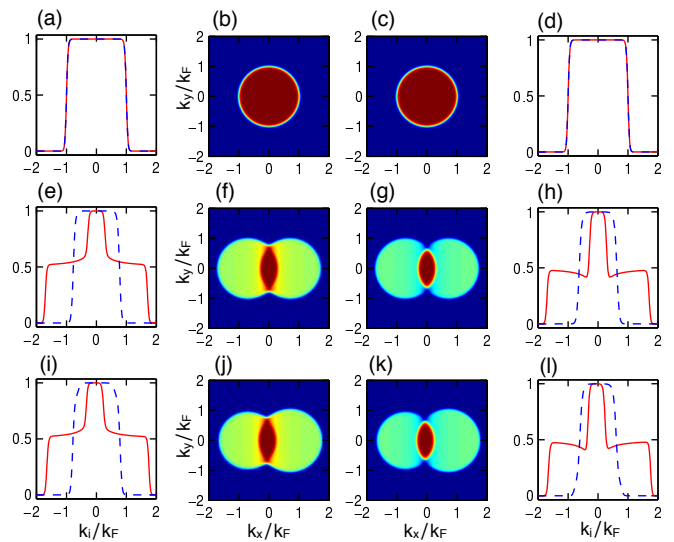}
\caption{
\label{fig:six}
(color online)
Finite temperature ($T/\epsilon_F = 0.05$) momentum distributions
$n_\uparrow ({\bf k})$ (two left-most columns)
and $n_\downarrow ({\bf k})$ (two right-most columns).
The parameters for (a)-(d) are ERD spin-orbit coupling
$v/v_F = 0$, Zeeman fields $h_y/\epsilon_F  = h_z/\epsilon_F = 0$,
chemical potential $\mu/\epsilon_F = 1.0$, and
induced polarization $P_{\rm ind} = 0$.
Similarly, the parameters for
(e)-(h) are $v/v_F = 0.71$, $h_y/\epsilon = 0$, $h_z/\epsilon_F = 0.1$,
$\mu/\epsilon_F = 0.495$, and $P_{\rm ind} = 0.121$; while for
(i)-(l) the parameters are
$v/v_F = 0.71$, $h_y/\epsilon_F = h_z/\epsilon_F = 0.1 $,
$\mu/\epsilon_F = 0.492$, and $P_{\rm ind} = 0.120$.
The blue dashed lines represent cuts of $n_s ({\bf k})$
along the ${\bf k} = (k_x = 0, k_y, k_z = 0)$ direction,
while the red solid lines represents cuts of $n_s ({\bf k})$
along ${\bf k} = (k_x, k_y = 0, k_z = 0)$.
}
\end{figure}

In Fig.~\ref{fig:six}, we show momentum distributions
$n_s ({\bf k})$ at $T/\epsilon_F = 0.05$ and
in the regime where the Fermi system is
largely degenerate, containing wide regions in momentum
space where $n_s ({\bf k}) \approx 1$.
The blue dashed lines represent cuts of $n_s ({\bf k})$
along the ${\bf k} = (k_x = 0, k_y, k_z = 0)$ direction,
while the red solid lines represents cuts of $n_s ({\bf k})$
along ${\bf k} = (k_x, k_y = 0, k_z = 0)$. The left-most
columns correspond to spin $\uparrow$, while the
right-most columns represent spin $\downarrow$.
For reference, we show in Figs.~\ref{fig:six}(a)-(d)
the case with zero spin-orbit coupling and without Zeeman fields,
corresponding to parameters
$v/v_F = 0$, $h_y/\epsilon_F = h_z/\epsilon_F = 0$
chemical potential $\mu = 1.0$.
In this case, the momentum distributions for the two spin components are
identical $n_\uparrow ({\bf k}) =  n_\downarrow ({\bf k})$,
meaning that the populations are balanced
with $n_{-} ({\bf k}) = 0$, and induced polarization is $P_{\rm ind} = 0$.
In Figs.~\ref{fig:six}(e)-(h), we show momentum distributions
for parameters
$v/v_F = 0.71$, $h_y/\epsilon_F = 0$, $h_z/\epsilon_F = 0.1$
and chemical potential $\mu = 0.495$.
Here, the momentum distributions acquire double {\it plateaux}
structures along the direction ${\bf k} = (k_x, k_y = 0, k_z = 0)$
due to the momentum shifts of the helicity bands
as shown in Fig.~\ref{fig:one}(a). Additionally, the momentum
distributions for different spin-components are no longer
identical, such that $n_\uparrow ({\bf k}) \ne n_\downarrow ({\bf k})$,
or $n_- ({\bf k}) \ne 0$, and the induced polarization
is non-zero taking the value $P_{\rm ind} = 0.121$.
In these two cases, parity is not violated and the momentum distributions
are even functions of momentum under inversion symmetry.
However, in Figs.~\ref{fig:six}(i)-(l), we show momentum
distributions for parameters
$v/v_F = 0.71$, $h_y/\epsilon_F = h_z/\epsilon_F = 0.1$,
and chemical potential $\mu = 0.492$,
in which case the double {\it plateaux} structures are still preserved,
population imbalance is present with $n_- ({\bf k}) \ne 0$ and induced
polarization $P_{\rm ind} = 0.120$. Most importantly parity
is weakly violated since $n_s (-{\bf k}) \ne n_s ({\bf k})$.

\begin{figure} [htb]
\includegraphics[width = 1.0\linewidth]{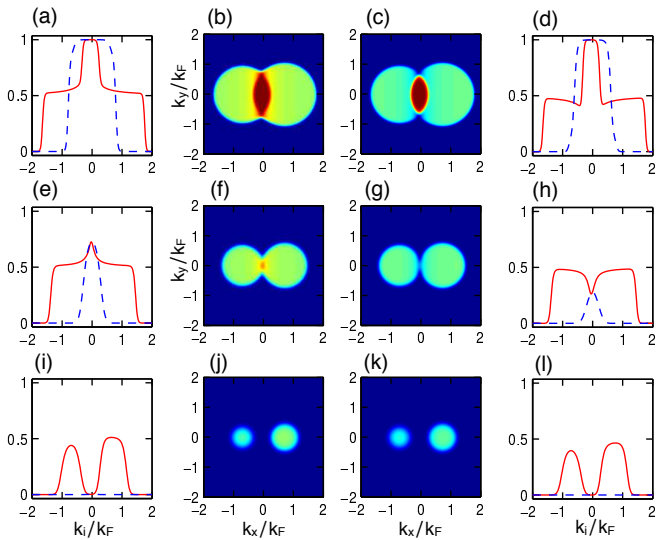}
\caption{
\label{fig:seven}
(color online)
Finite temperature ($T/\epsilon_F = 0.05$) momentum distributions
$n_\uparrow ({\bf k})$ (two left-most columns)
and $n_\downarrow ({\bf k})$ (two right-most columns).
All panels corresponds to values of ERD spin-orbit coupling
$v/v_F = 0.71$, and Zeeman fields $h_y/\epsilon_F  = h_z/\epsilon_F = 0.1$.
For (a)-(d) the chemical potential is $\mu/\epsilon_F = 0.492$, and the
induced polarization is $P_{\rm ind} = 0.120$.
Similarly, for (e)-(h) $\mu/\epsilon_F = 0$, and $P_{\rm ind} = 0.143$;
while for (i)-(l) the parameters are $\mu/\epsilon_F = -0.37$,
and $P_{\rm ind} = 0.105$.
The blue dashed lines represent cuts of $n_s ({\bf k})$
along the direction ${\bf k} = (k_x = 0, k_y, k_z = 0)$ direction,
while the red solid lines represents cuts of $n_s ({\bf k})$
along ${\bf k} = (k_x, k_y = 0, k_z = 0)$.
}
\end{figure}

In Fig.~\ref{fig:seven}, we show momentum distributions
$n_s ({\bf k})$ at $T/\epsilon_F = 0.05$ for parameters
$v/v_F = 0.71$,  $h_y/\epsilon_F = h_z/\epsilon_F = 0.1$,
and varying chemical potentials $\mu/\epsilon_F$.
We emphasize the regimes where parity is more strongly violated
leading to momentum distributions without inversion symmetry:
$n_s ({-\bf k}) \ne n_s ({\bf k})$.
The blue dashed lines represent cuts of $n_s ({\bf k})$
along the ${\bf k} = (k_x = 0, k_y, k_z = 0)$ direction,
while the red solid lines represents cuts of $n_s ({\bf k})$
along ${\bf k} = (k_x, k_y = 0, k_z = 0)$.
For reference, we show in Figs.~\ref{fig:seven}(a)-(d)
the case with $\mu/\epsilon_F = 0.492$ and $P_{\rm ind} = 0.120$,
where the Fermi system is still largely degenerate, containing
wide regions in momentum space with $n_s ({\bf k}) \approx 1$,
and at the same time parity is violated only weakly.
In Figs.~\ref{fig:seven}(e)-(h), we show momentum distributions
for $\mu/\epsilon_F = 0$ and $P_{\rm ind} = 0.143$,
while in Figs.~\ref{fig:seven}(i)-(l), we show momentum distributions
for $\mu/\epsilon_F = -0.37$ and $P_{\rm ind} = 0.105$.
In both cases, the Fermi system is no longer degenerate,
containing wide regions in momentum space
where $n_s ({\bf k}) \ll 1$.
In the last two cases the momentum distributions remain symmetric
upon reflection along the $k_y$ or $k_z$ directions, but
parity is more strongly violated leading to a highly
asymmetric momentum distributions along the
$k_x$ direction.

Having discussed the momentum distribution at low temperatures, we analyze
next the density of states of quantum degenerate fermions in the presence
of spin-orbit coupling and crossed Zeeman fields.

\section{Density of States}
\label{sec:density-of-states}

The density of states for spin $s$ can be written as
\begin{equation}
\rho_s (\omega)
=
\sum_{\bf k}  A_s (\bf k,\omega),
\end{equation}
in terms of the spectral function $A_s (\bf k,\omega).$
An analysis of the spin-dependent density of states is useful to
provide the frequency (energy) dependence of the spin-polarization
of the system.

\begin{figure} [htp]
\includegraphics[width = 1.0\linewidth]{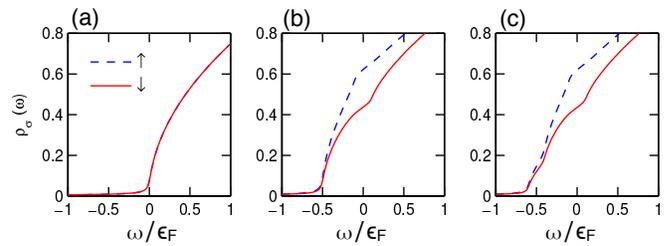}
\caption{
\label{fig:eight}
(color online)
Spin-dependent density of states $\rho_s (\omega)$
in units of $N/\epsilon_F$
for
(a) $v/v_F = 0$ and $h_z/\epsilon_F = h_y/\epsilon_F = 0$,
(b) $v/v_F = 0.71$ and $h_y/\epsilon_F = 0$ and $h_z/\epsilon_F = 0.1$,
and
(c) $v/v_F = 0.71$ and $h_z/\epsilon_F = h_y/\epsilon_F = 0.1$.
The blue dashed line corresponds to $\rho_\uparrow ({\omega})$
and the solid red line corresponds to $\rho_\downarrow ({\omega})$.
}
\end{figure}

In Fig.~\ref{fig:eight}, we show the density of states $\rho_s (\omega)$
for various parameters with an energy broadening
$\eta/\epsilon_F = 0.01$.
We show specifically the case of zero spin-orbit
coupling and Zeeman fields in Fig.~\ref{fig:eight}(a),
which has the characteristic square-root frequency dependence
for a three-dimensional system. In this case the system is not polarized
and the tails below the bottom of the energy band are due to the
finite energy broadening.
We show two other situations for comparison corresponding to cases which
are polarized with finite Zeeman fields, as well as with a finite
spin-orbit coupling. In Fig.~\ref{fig:eight}(b)-(c), we show that the
band edges shift to lower frequencies when the Zeeman and spin-orbit
fields are turned on. Furthermore, in Fig.~\ref{fig:eight}(b)
the spin-dependent density of states are shown for the case
where parity is not violated, corresponding to
$h_y/\epsilon_F =0$, $h_z/\epsilon_F = 0.1$, and $v/v_F = 0.71$;
and in Fig.~\ref{fig:eight}(c) the spin-dependent
density of states are shown for the case where parity is violated,
corresponding to $h_y/\epsilon_F = h_y/\epsilon_F =0.1$
and $v/v_F = 0.71$. Since momentum is integrated over,
there is no clear signature of parity violation in $\rho_s (\omega)$
as there is in momentum-resolved observables such as
the spectral density $A_s ({\bf k}, \omega )$,
momentum distribution $n_s ({\bf k})$, or helicity dispersions
$\epsilon_{\Uparrow} ({\bf k})$ and $\epsilon_{\Downarrow} ({\bf k})$
discussed earlier. However, we show the different spin-dependent
density of states for comparison. The kinks present in these figures
reflect the location of the maxima and minima of the helicity bands.

Having discussed the density of states, we present next an analysis of
thermodynamic properties.

\section{Thermodynamic Properties}
\label{sec:thermodynamic-properties}

From the energy spectrum of the Hamiltonian ${\bf H}_0$ defined
in Eq.~(\ref{eqn:hamiltonian-non-interacting}), we can obtain the
partition function as
\begin{equation}
\mathcal{Z}
=
{\rm Tr}
\left[
\exp(-{\bf H}_0/T)
\right],
\end{equation}
and the corresponding thermodynamic potential
\begin{equation}
\label{eq:thermodynamic-potential}
\Omega
=
-T \ln \mathcal{Z}.
\end{equation}
When no initial population imbalance is present,
the total number of particles is fixed by
\begin{equation}
\label{eqn:number}
N
=
-
\left(
\frac{\partial \Omega}{\partial \mu}
\right)_{T,V},
\end{equation}
which determines the chemical potential $\mu$.
In addition, because an external Zeeman field is
present, we can define the induced polarization
\begin{equation}
P_{\rm ind}
=
\frac{N_\uparrow - N_\downarrow}{N_\uparrow + N_\downarrow},
\end{equation}
where the number of particles in spin-state $s$ is
\begin{equation}
N_s
=
\sum_{\bf k}
\int d\omega
n_F(\omega) A_s ({\bf k},\omega).
\end{equation}

Next, we begin our analysis of thermodynamic variables
that could be measured using the techniques already developed
for ultra-cold fermions~\cite{salomon-2010, zwierlein-2012a}
in the absence of spin-orbit coupling. In the discussion
that follows, we will cover the pressure, the chemical potential,
the isothermal compressibility and the induced magnetization
(spin-polarization).

\section{Pressure and Entropy}
\label{sec:pressure-entropy}

The pressure of ultra-cold fermions in the presence of spin-orbit
coupling and crossed Zeeman fields is
\begin{equation}
P
=
-\frac{\Omega}{V}
=
-\frac{T}{V}
\sum_{{\bf k},\alpha}
\ln
\left(
1 + e^{-\xi_\alpha ({\bf k})/T}
\right),
\end{equation}
where $\Omega$ is thermodynamic potential, and
$\alpha = ( \Uparrow,\Downarrow )$ is the
helicity spin index.

\begin{figure} [htb]
\includegraphics[width = 1.0\linewidth]{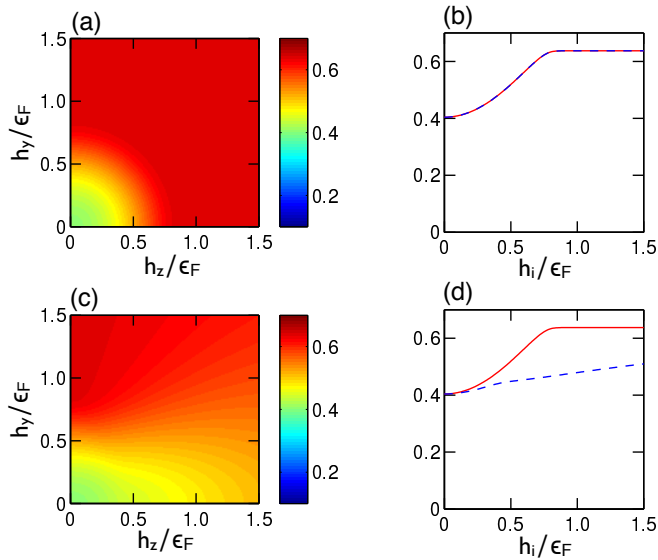}
\caption{
\label{fig:nine}
(color online)
Pressure $PV / N\epsilon_F$ as a function of $h_y/\epsilon_F$ and
$h_z/\epsilon_F$ at finite temperature ($T/\epsilon_F = 0.05$)
for (a)-(b) $v/v_F = 0$ and (c)-(d) $v/v_F = 0.71$.
The red solid line represents
$PV/N\epsilon_{F}$ as a function of $h_y/\epsilon_F$
at $h_z/\epsilon_F  = 0$, and the blue dashed line
represents $PV/N\epsilon_{F}$ as a function of $h_z/\epsilon_F$ at
$ h_y/\epsilon_F = 0$.
}
\end{figure}

In Fig.~\ref{fig:nine}, we show the scaled pressure
$PV/(N \epsilon_F)$ as a function of the Zeeman fields
$h_y/\epsilon_F$ and $h_z/\epsilon_F$ for fixed spin-orbit coupling $v/v_F$.
Notice that in the absence of spin-orbit and Zeeman fields
the pressure reduces to the standard results of
an non-interacting Fermi gas $PV/(N \epsilon_F) = 2/5$.
The pressure is an even function of both $h_y$ and $h_z$
and is shown for a range of $h_y$ and $h_z$ varying from zero
to $1.5 \epsilon_F$.  The spin-orbit coupling for
Figs.~\ref{fig:nine}(a)-(b) is $v/v_F = 0$, and the pressure is
completely isotropic in the $h_y$-$h_z$ plane,
since the helicity bands $\epsilon_{\alpha} ({\bf k})$
are also isotropic in $h_y$-$h_z$ plane and
parity is preserved.
However, in Fig.~\ref{fig:nine}(c)-(d), where $v/v_F = 0.71$,
the pressure is anisotropic in the $h_y$-$h_z$ plane,
since the helicity bands $\epsilon_{\alpha} ({\bf k})$
are now anisotropic in the $h_y$-$h_z$ plane and are
not invariant under parity.

We note in passing that the entropy $S$ of the system
can be easily extracted, by rewriting the thermodynamic potential as
$
\Omega
=
T \sum_{\bf k}
\ln
\left[
1 - n_F (\xi_{\alpha} ({\bf k}))
\right],
$
when expressed in terms of the Fermi function $n_F (\xi_{\alpha} ({\bf k}))$.
Using the relation
$
S
=
-
\left(
\partial \Omega / \partial T
\right)_{V, \mu},
$
leads to the final result ($k_B = 1$)
\begin{equation}
S
=
-
\sum_{{\bf k} \alpha, m = \pm}
\left\{
n_F (m \xi_{\alpha} ({\bf k}))
\ln
\left[
n_F (m \xi_{\alpha} ({\bf k}))
\right]
\right\},
\end{equation}
which is nothing but the entropy of a non-interacting Fermi gas in the
presence of spin-orbit coupling and crossed Zeeman fields. We will not
show plots of the entropy, or of the specific heat, which can also be
easily obtained, but rather discuss next the chemical potential and
its dependence on the crossed Zeeman fields.

\section{Chemical Potential}
\label{sec:chemical-potential}

\begin{figure} [htb]
\includegraphics[width = 1.0\linewidth]{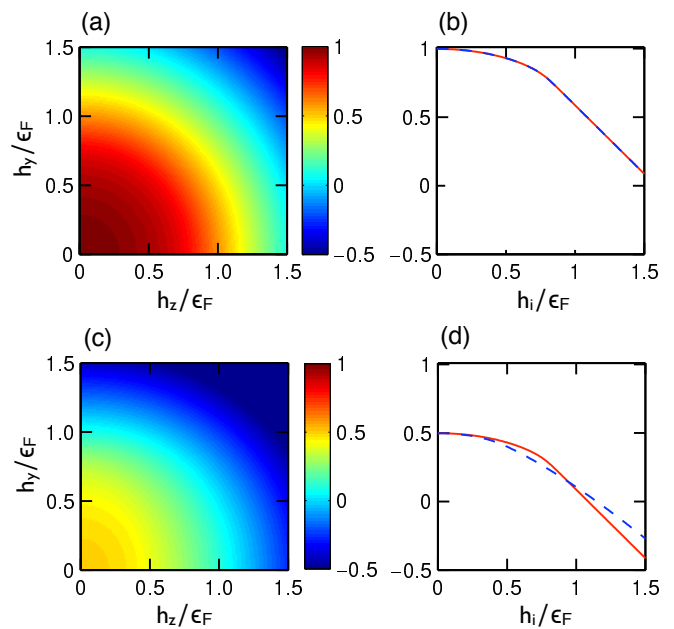}
\caption{
\label{fig:ten}
(color online)
Finite temperature ($T/\epsilon_F = 0.05$) chemical potential
$\mu / \epsilon_F$ as a function of $h_z$ and $h_y$
is shown in (a)-(b) $v/v_F = 0$ and in (c)-(d) for $v/v_F = 0.71$.
The red solid line represents $\mu (h_y, h_z = 0)/\epsilon_{F}$
and the blue dashed line represents $\mu (h_y = 0,h_z)/\epsilon_{F}$.
}
\end{figure}

The chemical potential $\mu$ in the Grand-canonical ensemble is determined
by fixing the average number of particles given in Eq.~(\ref{eqn:number}),
which can be rewritten as
\begin{equation}
\label{eqn:number-total}
N
=
\sum_{\bf k} n_+ ({\bf k}),
\end{equation}
where $n_+ ({\bf k})$ is the total momentum distribution
defined in Eq.~(\ref{eqn:momentum-distribution-sum}).
The behavior of $\mu$ as a function of the Zeeman fields
$h_y$ and $h_z$ is shown in Fig.~\ref{fig:ten},
which uses the fact that $\mu$ is an even function of these variables.
The range of the Zeeman fields is also from $0$ to $1.5 \epsilon_F$.
The case where parity is preserved is shown if Fig.~\ref{fig:ten}(a) and (b),
corresponding to $v/v_F = 0$, such that the chemical potential is
isotropic in the $h_y$-$h_z$ plane. Similarly,
in Fig.~\ref{fig:ten}(c) and (d), we show the case corresponding
to $v/v_F = 0.71$, where parity is violated for any finite
Zeeman component $h_y$. As a result the chemical potential
$\mu/\epsilon_F$ is anisotropic in $h_y$-$h_z$ plane,
because the helicity bands are neither even nor odd in momentum
space. This demonstrates that the anisotropy of $\mu$ is a measure of
parity violation. Another quantity that can be easily measured and
that reveals a similar effect is the isothermal compressibility
to be discussed next.

\section{Isothermal Compressibility}
\label{sec:isothermal-compressibility}

The isothermal compressibility can be obtained from the knowledge
of the pressure, as is defined as
$
\kappa_T
=
- (1/V)
(\partial P/\partial V)_{T,N}.
$
But in the Grand-canonical ensemble, where we need to fix
the average number of particles, the isothermal compressibility
can be directly written as
\begin{equation}
N^2 \kappa_T
=
\left(
\frac{\partial N}{\partial \mu}
\right)_{T,V}.
\end{equation}

Using the relation defined in Eq.~(\ref{eqn:number-total})
and noticing that the partial derivative
$
\left(
\partial n_F(x) / \partial x
\right)_{T,V}
=
-(1/T) n_F(x) n_F(-x),
$
is directly related to the Fermi function $n_F (x)$,
it is possible to write the expression for the isothermal compressibility
as
\begin{equation}
N^2 \kappa_T
=
\frac{1}{T}\sum_{{\bf k}, \alpha}
\left[
n_F(\xi_{\alpha} ({\bf k}))  n_F(-\xi_{\alpha} ({\bf k}))
\right].
\end{equation}

The experimental extraction of the isothermal compressibility
from measurements of density fluctuations was suggested
theoretically several years ago both in harmonically confined
systems~\cite{iskin-2005} and optical lattices~\cite{iskin-2006},
and early improvements in the detection schemes of density
fluctuations~\cite{bloch-2005,bouchole-2006,steinhauer-2010}
became sufficiently sensitive to extract this information
from experimental data.
In a recent experiment~\cite{ketterle-2011}
using laser speckles, the isothermal compressibility
and the spin susceptibility were measured as
a function of interaction parameter via the fluctuation dissipation
theorem throughout the evolution from BCS
to BEC superfluidity in balanced Fermi systems.

The atomic compressibility can be measured using
the fluctuation-dissipation theorem relating the
fluctuation in the density of particles $n = N/V$, where
$N = \langle {\hat N} \rangle$ is the average number of particles,
and ${\hat N}$ is particle-number operator.
The relation between the isothermal compressibility $\kappa_T$
and particle-number
(density) fluctuations is given by the relation:
\begin{equation}
\kappa_T
=
\frac{V}{T}
\frac{ \langle {\hat N}^2 \rangle - \langle {\hat N} \rangle^2}
{\langle {\hat N} \rangle^2}.
\end{equation}

We see no major technical impediment to use techniques
that are sensitive to spin-dependent density fluctuations
in population imbalanced Fermi-Fermi mixtures with equal
masses~\cite{ketterle-2006,hulet-2006}, whether
the imbalance is created initially via radio-frequency fields
or via artificial spin-orbit and Zeeman fields. Such analysis
was shown to be theoretically possible
even for Fermi mixtures with unequal masses~\cite{seo-2011a,seo-2011b},
and preliminary experimental results for
these systems~\cite{grimm-2010,grimm-2011}
seem to indicate that indeed the compressibility and spin susceptibility
matrix elements can be directly extracted from the local density
and density fluctuation profiles.

\begin{figure} [htb]
\includegraphics[width = 1.0\linewidth]{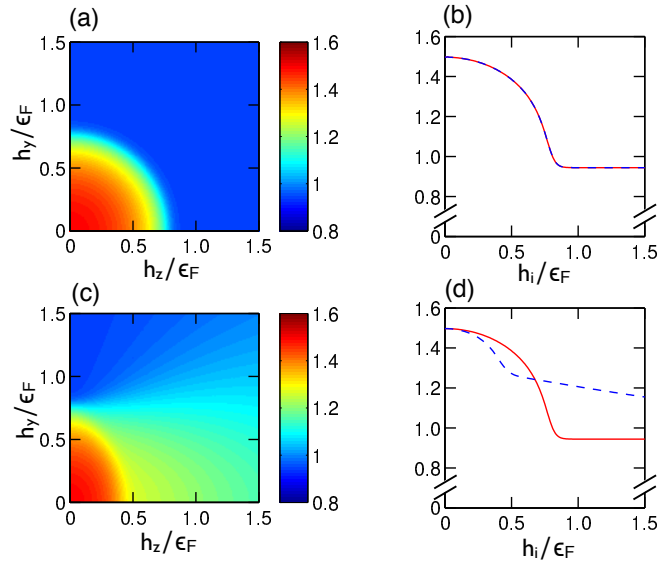}
\caption{
\label{fig:eleven}
(color online)
Finite temperature ($T/\epsilon_F = 0.05$) isothermal compressibility
$N^2\kappa_T$ in units of $N/\epsilon_F$
as a function of $h_y$ and $h_z$
is shown in (a)-(b) for $v/v_F = 0$ and in (c)-(d) for $v/v_F = 0.71$.
The red solid line represents
$N^2 \kappa_{T} (h_y, h_z = 0)$
and the blue dashed line represents
$N^2 \kappa_{T} (h_y = 0,h_z)$.
}
\end{figure}

Thus, we show in Fig.~\ref{fig:eleven} the isothermal
compressibility $\kappa_T$ as a function of both $h_y$ and $h_z$
for a range of $h_y$ and $h_z$ varying from zero to $1.5 \epsilon_F$,
at fixed temperature $(T = 0.05\epsilon_F)$.
The spin-orbit coupling for
Figs.~\ref{fig:eleven}(a)-(b) is $v/v_F = 0$, and the compressibility
is completely isotropic, as parity is preserved. However,
in Fig.~\ref{fig:eleven}(c)-(d), where $v/v_F = 0.71$,
parity is broken for $h_y \ne 0$.
This parity breaking is reflected in the helicity bands as discussed
earlier, and manifests itself in the behavior of the
compressibility versus $(h_y, h_z)$ via the anisotropy
revealed in Fig.~\ref{fig:eleven}(c)-(d). Again such anisotropy is
a reflection of the parity violation caused by the simultaneous
presence of the ERD spin-orbit field and $h_y$. Another important
property that can be measured is the spin polarization as a function
of the Zeeman fields $h_y, h_z$ for fixed spin-orbit coupling, which
is discussed next.

\section{Induced Spin Polarization}
\label{sec:induced-polarization}

We can also analyze the induced spin polarization in the presence
of crossed Zeeman fields and spin-orbit coupling. The spin polarization
along the $i^{th}$ direction $(i = x, y, z)$ is given by the expectation
value
\begin{equation}
\langle
{\hat S}_i
\rangle
=
\frac{1}{2}
\sum_{\bf k}
\langle
\Psi^\dagger ({\bf k})
{\mathbf \sigma}_i
\Psi ({\bf k})
\rangle.
\end{equation}
Such general expression can be particularized for each component.
For instance, the expectation value
\begin{equation}
\langle {\hat S}_x \rangle
=
\frac{1}{2} \sum_{\bf k}
\left(
\langle
c_{{\bf k} \uparrow}^\dagger c_{{\bf k} \downarrow}
\rangle
+
\langle
c_{{\bf k} \downarrow}^\dagger c_{{\bf k} \uparrow}
\rangle
\right).
\end{equation}
expressed in the original spin basis, can be
rewritten in the helicity basis as
\begin{equation}
\langle {\hat S}_x \rangle
=
-\frac{1}{2} \sum_{\bf k}
u_{\bf k}( v_{\bf k} + v_{\bf k}^* )
\left[
n_F (\xi_\Uparrow ({\bf k}))
-
n_F (\xi_\Downarrow ({\bf k}))
\right].
\end{equation}
Analogously the expectation value of the spin-operator
${\hat S}_y$ in the original spin basis is
\begin{equation}
\langle {\hat S}_y \rangle
=
-\frac{i}{2} \sum_{\bf k}
\left[
\langle
c_{{\bf k} \uparrow}^\dagger c_{{\bf k} \downarrow}
\rangle
-
\langle
c_{{\bf k} \downarrow}^\dagger c_{{\bf k} \uparrow}
\rangle
\right]
\end{equation}
in the original spin basis can be written as
\begin{equation}
\langle {\hat S}_y \rangle
=
-\frac{i}{2} \sum_{\bf k}
u_{\bf k}( v_{\bf k}- v_{\bf k}^* )
\left[
n_F ( \xi_\Uparrow ({\bf k}) )
-
n_F ( \xi_\Downarrow ( {\bf k}) )
\right]
\end{equation}
in the helicity basis.
Lastly, the expectation value of ${\hat S}_z$ in the
original spin basis is
\begin{equation}
\langle {\hat S}_z \rangle
=
\frac{1}{2}
\sum_{\bf k}
\left(
\langle
c_{{\bf k} \uparrow}^\dagger c_{{\bf k} \uparrow}
\rangle
-
\langle
c_{{\bf k} \downarrow}^\dagger c_{{\bf k} \downarrow}
\rangle
\right),
\end{equation}
which can be rewritten in the helicity basis as
\begin{equation}
\langle {\hat S}_z \rangle
=
\frac{1}{2}
\sum_{\bf k}
(
u_{\bf k}^2 - |v_{\bf k}|^2
)
\left[
 n_{F} ( \xi_{\Uparrow} ({\bf k}) )
-
n_{F} ( \xi_{\Downarrow}({\bf k}) )
\right].
\end{equation}

Finally, the averages $\langle {\hat S}_x \rangle$ and
$\langle {\hat S}_y \rangle$ can be expressed as real and
imaginary parts of the transverse spin-polarization
$
\langle {\hat S}_\perp \rangle
=
\langle {\hat S}_x \rangle
-
i \langle {\hat S}_y \rangle,
$
defined this way to be compatible with
the definition of
$h_{\perp} ({\bf k}) = h_x ({\bf k}) - i h_y ({\bf k})$.
The transverse spin-polarization takes
the final form
\begin{equation}
\langle {\hat S}_{\perp} \rangle
=
\frac{1}{2}
\sum_{\bf k}
\frac{h_\perp({\bf k})}{\lvert h_{\rm eff}({\bf k})\rvert}
\left[
n_F (\xi_\Uparrow ({\bf k}) )
-
n_F (\xi_\Downarrow ({\bf k}) )
\right].
\end{equation}
Correspondingly the longitudinal spin polarization can
be written as
\begin{equation}
\langle {\hat S}_z \rangle
=
\frac{1}{2}
\sum_{\bf k}
\frac{h_\parallel}{\lvert h_{\rm eff}({\bf k})\rvert}
\left[
n_F (\xi_\Uparrow ({\bf k}) )
-
n_F (\xi_\Downarrow ({\bf k}) )
\right],
\end{equation}
which is directly related to the induced population
imbalance $P_{\rm ind}$ by the expression
\begin{equation}
P_{\rm ind}
=
\frac{2 \langle {\hat S}_z \rangle}{N}
\end{equation}
where
$
N
=
N_{\uparrow}
+
N_{\downarrow}
$
is the total number of particles, as defined earlier.

\begin{figure} [htb]
\includegraphics[width = 1.0\linewidth]{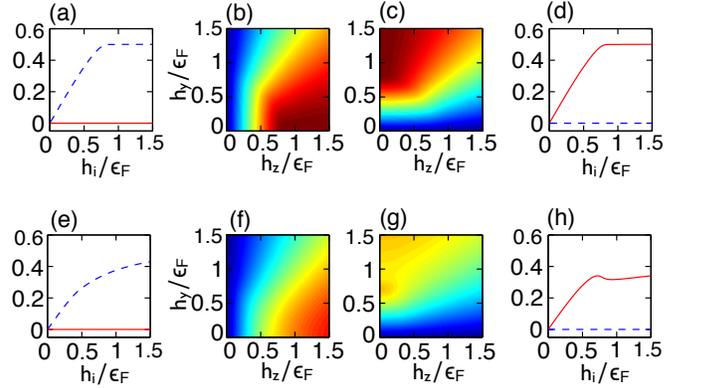}
\caption{
\label{fig:twelve}
(color online)
Finite temperature ($T/\epsilon_F = 0.05$)
induced spin polarizations per particle
$\langle S_z \rangle/N$ (left two columns)
and $\langle S_y \rangle/N$ (right two columns) are shown in
(a)-(d) for $v/v_F = 0$ and (e)-(h) for $v/v_F = 0.71$.
In (a) and (e), the red solid line represents
$\langle S_z \rangle/N$ as a function of $h_y/\epsilon_F$
at $h_z/\epsilon_F = 0$, and the blue dashed line represents
$\langle S_z \rangle/N$ as a function of $h_z/\epsilon_F$ at
$h_y/\epsilon = 0$.
In (d) and (h), the red solid line represents
$\langle S_y \rangle/N$ as a function of $h_y/\epsilon_F$ at
$h_z/\epsilon_F = 0$, and the blue dashed line represents
$\langle S_y \rangle/N$ as a function of $h_z/\epsilon_F$ at
$h_y/\epsilon_F = 0$.
}
\end{figure}

For ERD spin-orbit coupling
with field ${\bf h}_{ERD} = v k_x {\hat {\bf y}}$, the transverse
field $h_{\perp} ({\bf k})$ has only the y-component. This means that
$\langle {\hat S}_x \rangle$ is identically zero for any value
of $h_y$ and $h_z$ for any value of $v \ne 0$,
given that $h_x ({\bf k}) = 0$.
However, $\langle {\hat S}_y \rangle$ is not identically zero for
the ERD case above, unless parity is preserved in the
helicity bands $\xi_{\alpha} ({\bf k})$, which means $h_y = 0$.
For any finite value of $h_y$, $\langle {\hat S}_y \rangle$
is non-zero. This behavior is revealed in Fig.~\ref{fig:twelve},
where the expectation values $\langle {\hat S}_y \rangle$
and $\langle {\hat S}_z \rangle$ of spin polarization are shown as
a function of $(h_y/\epsilon_F, h_z/\epsilon_F)$.
In particular, we show in Fig.~\ref{fig:twelve}
the finite temperature ($T/\epsilon_F = 0.05$)
induced spin polarizations per particle
$\langle S_z \rangle/N$ (two left-most columns)
and $\langle S_y \rangle/N$ (two right-most columns)
for ERD spin-orbit parameter $v/v_F = 0$ from (a) through (d),
and for $v/v_F = 0.71$ from (e) through (h).
In (a) and (e), the red solid line represents
$\langle S_z \rangle/N$ as a function of $h_y/\epsilon_F$ at
$h_z/\epsilon_F = 0$, and
the blue dashed line represents
$\langle S_z \rangle/N$ as a function of
$h_z\epsilon_F$ at $h_y/\epsilon_F = 0$.
In (d) and (h), the red solid line represents
$\langle S_y \rangle/N$ as a function of $h_y/\epsilon_F$ at
$h_z/\epsilon_F = 0$, and
the blue dashed line represents
$\langle S_y \rangle/N$ as a function of $h_z/\epsilon_F$ at
$h_y/\epsilon_F = 0$.

Having analyzed several thermodynamic properties for
non-interacting quantum degenerate ultra-cold fermions, which
already present some fundamental non-trivial properties such
as the violation of parity, we discuss next the effects of
interactions and how parity violation affects the pairing
temperature of such fermions and the superfluid order parameter.

\section{Effects of Interactions}
\label{sec:effects-of-interactions}

In this section, we analyze briefly the effects of interactions
in the presence of spin-orbit and crossed Zeeman fields, with
focus on the dependence of the pairing temperature with respect
to the interaction parameter for given spin-orbit and Zeeman fields.
We also present a short discussion about the effects of parity violation
on the superfluid order parameter.

The interaction Hamiltonian is
\begin{equation}
\label{eqn:hamiltonian-interaction}
{\cal H}_I
=
-g
\int d{\bf r}
\psi^{\dagger}_{\uparrow} ({\bf r})
\psi^{\dagger}_{\downarrow} ({\bf r})
\psi_{\downarrow} ({\bf r})
\psi_{\uparrow} ({\bf r}),
\end{equation}
where $g$ represents the strength of the contact interaction.
Only $s$-wave scattering is considered in regards to the
original spin states $\uparrow$ and $\downarrow$.
Converting the interaction term into momentum space leads to

\begin{equation}
{\cal H}_I
=
-g
\sum_{\bf q}
b^\dagger ({\bf q})
b ({\bf q}),
\end{equation}
where the pair creation operator with center
of mass momentum ${\bf q}$ is
$
b^\dagger ({\bf q})
=
\sum_{{\bf k}}
\psi^\dagger_{\uparrow} ( {\bf k} + {\bf q}/2 )
\psi^\dagger_{\downarrow} (-{\bf k} + {\bf q}/2 ),
$
%
%
and $g$ can be expressed in terms of the scattering length
through the Lippman-Schwinger relation
\begin{equation}
\label{eqn:lippman-schwinger-relation}
\frac{V}{g}
=
-\frac{V m}{4\pi a_s}
+
\sum_{\bf k} \frac{1}{2\epsilon_{\bf k}}.
\end{equation}

The interaction Hamiltonian ${\cal H}_{I}$
can be written in the helicity basis as
\begin{equation}
{\widetilde {\cal H}}_{I}
=
-g
\sum_{{\bf q} \alpha \beta \gamma \delta}
B^\dagger_{\alpha \beta} ({\bf q})
B_{\gamma \delta} ({\bf q}),
\end{equation}
where the indices $\alpha, \beta, \gamma, \delta$ cover
$\Uparrow$ and $\Downarrow$ states. Pairing is now described
by the operator
\begin{equation}
\label{eqn:pairing-operator-helicity-basis}
B_{\alpha \beta} ({\bf q})
=
\sum_{\bf k}
\Lambda_{\alpha \beta}
({\bf k}_1, {\bf k}_2)
\Phi_{\alpha} ({\bf k}_1 )
\Phi_{\beta} ({\bf k}_2)
\end{equation}
and its Hermitian conjugate, with momentum indices
${\bf k}_1 = {\bf k} + {\bf q}/2$
and
${\bf k}_2 = -{\bf k} + {\bf q}/2$.
The matrix
$
\Lambda_{\alpha \beta}
({\bf k} + {\bf q}/2, -{\bf k} + {\bf q}/2)
$
is directly related to the matrix elements of the momentum
dependent SU(2) rotation
into the helicity basis,
and reveals that the center of mass momentum
${\bf k}_1 + {\bf k}_2 = {\bf q}$ and
the relative momentum ${\bf k}_1 - {\bf k}_2 = 2{\bf k}$
are coupled and no longer independent.

%
%

\section{Tensor Order Parameter}
\label{sec:tensor-order-parameter}

From Eq.~(\ref{eqn:pairing-operator-helicity-basis}) it is
clear that pairing between fermions of momenta
${\bf k}_1$ and ${\bf k}_2$ can occur within the
same helicity band (intra-helicity pairing)
or between two different helicity bands
(inter-helicity pairing). For pairing at ${\bf q} = 0$,
the order parameter for superfluidity is the tensor
$
\Delta_{\alpha \beta} ({\bf k})
=
\Delta_0
\Lambda_{\alpha \beta} ({\bf k}, -{\bf k}),
$
where
$
\Delta_0
=
- g
\sum_{\gamma \delta}
\langle
B_{\gamma \delta} ({\bf 0})
\rangle,
$
leading to
components:
\begin{equation}
\Delta_{\Uparrow \Uparrow} ({\bf k})
=
\Delta_0
\left(
u_{\bf k} v_{-\bf k}
-
v_{\bf k} u_{-\bf k}
\right)
\end{equation}
for total helicity projection
$\lambda = +1$;
\begin{equation}
\begin{array}{l}
\Delta_{\Uparrow \Downarrow} ({\bf k})
=
-
\Delta_0
\left(
u_{\bf k} u_{-\bf k}
+
v_{\bf k} v_{-\bf k}^*
\right)
\\\\
\Delta_{\Downarrow \Uparrow} ({\bf k})
=
\Delta_0
\left(
u_{\bf k} u_{-\bf k}
+
v_{\bf k}^* v_{-\bf k}
\right)
\end{array}
\end{equation}
for total helicity projection $\lambda = 0$; and
\begin{equation}
\Delta_{\Downarrow \Downarrow} ({\bf k})
=
\Delta_0
\left(
u_{\bf k} v_{-\bf k}^*
-
v_{\bf k}^* u_{-\bf k}
\right)
\end{equation}
for total helicity projection $\lambda = -1$.

It is very important to emphasize that for non-zero
spin-orbit coupling and crossed Zeeman fields $h_y$
and $h_z$, the order parameter tensor $\Delta_{\alpha \beta} ({\bf k})$
does not have well defined parity.
For instance, while $\Delta_{\Uparrow \Uparrow} ({\bf k})$ and
$\Delta_{\Downarrow \Downarrow} ({\bf k})$ have odd parity,
the matrix elements $\Delta_{\Uparrow \Downarrow} ({\bf k})$ and
$\Delta_{\Downarrow \Uparrow} ({\bf k})$ do not have well defined parity.
However, we may still define singlet
and triplet sectors for the helicity basis, such that the singlet
sector
$
\Delta_{S,0} ({\bf k})
=
\left[
\Delta_{\Uparrow \Downarrow} ({\bf k})
-
\Delta_{\Downarrow \Uparrow} ({\bf k})
\right]/2
$
has even parity and the triplet
sector defined by the components
$\Delta_{\Uparrow \Uparrow} ({\bf k})$,
$
\Delta_{T,0} ({\bf k})
=
\left[
\Delta_{\Uparrow \Downarrow} ({\bf k})
+
\Delta_{\Downarrow \Uparrow} ({\bf k})
\right]/2
$
and
$\Delta_{\Downarrow \Downarrow} ({\bf k})$
have odd parity for any value of the
ERD spin-orbit coupling $v/v_F$
and crossed Zeeman fields $h_y/\epsilon_F$ and $h_z/\epsilon_F$.
The preservation of parity in the singlet and triplet sectors
is also true for the Rashba-only (RO) case, but the
order parameter breaks time-reversal symmetry.

Within the mean field approximation, the Hamiltonian
matrix in the helicity basis is
\begin{equation}
\label{eqn:mean-field-hamiltonian-matrix}
{\widetilde {\bf H}}_{\rm MF} ({\bf k})
=
\begin{pmatrix}
\xi_{{\bf k}\Uparrow} & 0 &
\Delta_{\Uparrow\Uparrow}({\bf k}) & \Delta_{\Uparrow\Downarrow}({\bf k}) \\
0 & \xi_{{\bf k} \Downarrow}  &
\Delta_{\Downarrow \Uparrow}({\bf k}) & \Delta_{\Downarrow\Downarrow}({\bf k}) \\
\Delta_{\Uparrow\Uparrow}^* ({\bf k}) & \Delta_{\Downarrow\Uparrow}^*({\bf k})
& -\xi_{-{\bf k} \Uparrow} & 0 \\
\Delta_{\Uparrow\Downarrow}^*({\bf k}) & \Delta_{\Downarrow\Downarrow}^*({\bf k})
& 0 & -\xi_{-{\bf k}\Downarrow}
\end{pmatrix}.
\end{equation}
This Hamiltonian matrix is traceless, therefore
the sum of its eigenvalues is zero,  however the eigenvalues of
$\widetilde {\bf H}_{\rm MF} ({\bf k})$ are not invariant
under parity. By labeling the eigenvalues as $E_1 ({\bf k})$,
$E_2 ({\bf k})$, $E_3 ({\bf k})$ and $E_4 ({\bf k})$ in decreasing order
of energy, and using the tracelessness condition then the sum
$
E_3 ({\bf k}) + E_4 ({\bf k})
=
-
\left[
E_1 ({\bf k}) + E_2 ({\bf k})
\right]
$
but each eigenvalue $E_i ({\bf k})$ does not a well
defined parity. Typically these eigenvalues are even in momentum space,
but not here because the parity violation induced by the simultaneous
presence of the crossed Zeeman fields and the spin-orbit coupling,
thus, in the present case $E_i (-{\bf k}) \ne E_i ({\bf k})$.
However a generalized particle-hole symmetry applies leading to
$E_2 ({\bf k}) = - E_3 (-{\bf k})$ and
$E_1 ({\bf k}) = - E_4 (-{\bf k}).$

The eigenvalues in this parity violating case can be obtained
analytically for any mixture of Rashba and Dresselhaus terms from the
determinant
$
{\rm Det}
\left[
\omega {\bf 1} -  \widetilde{\bf H}_{\rm MF} ({\bf k})
\right],
$
which leads to the characteristic quartic equation
\begin{equation}
\omega^4
+
a_3 ({\bf k}) \omega^3
+
a_2 ({\bf k}) \omega^2
+
a_1 ({\bf k}) \omega
+
a_0 ({\bf k})
=
0
\end{equation}
for each momentum ${\bf k}$.
Here, the coefficient of the cubic term
$a_3 ({\bf k}) = -\sum_i E_i ({\bf k})$
is the sum of the eigenvalues $E_i ({\bf k})$ of the
the Hamiltonian matrix $\widetilde{\bf H}_{\rm MF} ({\bf k})$
and therefore vanishes. The coefficient of the
quadratic term is
$a_2 ({\bf k}) = \sum_{i < j} E_i ({\bf k})  E_j ({\bf k})$,
while the coefficient of the linear
term is
$
a_1 ({\bf k})
=
\sum_{i \ne j \ne \ell}
E_i ({\bf k})  E_j ({\bf k}) E_\ell ({\bf k}).
$
The last coefficient is just the product of the four eigenvalues
leading to
$a_0 = E_1({\bf k}) E_2 ({\bf k}) E_3 ({\bf k}) E_4 ({\bf k})$.

In the particular case of ERD spin-orbit coupling with crossed Zeeman
fields, the coefficients become $a_3 = 0$, and
the coefficient of the quadratic term takes the form
\begin{equation}
a_2 =
- 2
\left(
{\widetilde K}_+^2({\bf k})
+ |\Delta_0|^2 + |v k_x|^2 + |h_y|^2 + |h_z|^2
\right),
\end{equation}
while the coefficient of the linear term is
\begin{equation}
a_1 =
-8  {\widetilde K}_+({\bf k}) ( v k_x ) h_y,
\end{equation}
and lastly the coefficient of the zero-th order
term is
\begin{equation}
a_0 =
\xi_{\Uparrow}({\bf k})
\xi_{\Downarrow}({\bf k})
\xi_{\Uparrow}({-\bf k})
\xi_{\Downarrow}(-{\bf k})
+
|\Delta_0|^2 \alpha_0^2 ({\bf k})
\end{equation}
where
$
\alpha_0^2 ({\bf k})
=
\left(
 2 {\widetilde K}_+^2({\bf k})
+ |\Delta_0|^2 + h_0^2 ({\bf k})
\right)
$
with
$
h_0^2 ({\bf k})
=
2|v k_x|^2 -2 |h_y|^2 -2 |h_z|^2.
$
Here, ${\widetilde K}_+$ has the same definition
used in the paragraph that follows
Eq.~(\ref{eqn:hamiltonian-matrix-non-interacting}),
and is a measure of the kinetic energy
with respect to the chemical potential.

Even in this simpler case of ERD spin-orbit coupling,
the precise analytical form of the eigenvalues in the presence
of crossed Zeeman fields is quite cumbersome, and
we do not list them here explicitly. Rather, we discuss next
the consequences of parity violation on the pairing temperature
of ultra-cold fermions in the presence of spin-orbit and crossed
Zeeman fields.

\section{Pairing Temperature}
\label{sec:pairing-temperature}

From the excitation spectrum discussed above, we obtain
the corresponding thermodynamic potential as
\begin{equation}
\Omega_{\rm MF}
=
V \frac{|\Delta_0|^2}{g}
-
\frac{T}{2}
\sum_{{\bf k}, j=1,4}
\ln
\left(
1 + e^{-E_j({\bf k})/T}
\right)
+
\sum_{{\bf k}}
\widetilde K_+ ({\bf k}),
\end{equation}
from which the order parameter equation is determined
via the minimization of
$\Omega_{\rm MF}$ with respect to $\vert \Delta_0 \vert^2$,
leading to
\begin{equation}
\label{eqn:order-parameter}
\frac{V}{g}
=
-\frac{1}{2}
\sum_{{\bf k}, j}
n_F \left(E_j ({\bf k}) \right)
\frac{\partial E_j ({\bf k})}{\partial \vert \Delta_0 \vert^2},
\end{equation}
where
$
n_F \left(  E_j (\mathbf{k})  \right)
=
1/(\exp\left( E_j ({\bf k})/T\right) + 1)
$
is the Fermi function for energy $E_j ({\bf k})$.
The contact interaction $g$ can be elliminated
in favor of the scattering length $a_s$ via the
Lippman-Schwinger relation defined in
Eq.~(\ref{eqn:lippman-schwinger-relation}).

The total number of particles $N = N_\uparrow + N_\downarrow$ is
defined from the thermodynamic relation
$
N
=
-
\left(
\partial \Omega_{\rm MF}
/
\partial \mu
\right)_{T, V},
$
and leads to the corresponding
number equation
\begin{equation}
\label{eqn:number-saddle-point}
N_{\rm MF}
=
\sum_{\bf k}
\left[
1 -
\frac{1}{2}\sum_j n_F \left( E_j ({\bf k}) \right)
\frac{\partial E_j ({\bf k})}{\partial \mu}
\right],
\end{equation}
since the system is assumed to have no initial population imbalance.

The self-consistent solutions of Eq.~(\ref{eqn:order-parameter})
and~(\ref{eqn:number-saddle-point}) guarantee the existence of mean field
solutions for the order parameter amplitude $\vert \Delta_0 \vert$
and the chemical potential $\mu$ as a function of the Zeeman fields
$h_y$ and $h_z$, the spin-orbit coupling $v$ and scattering length $a_s$.
However, the thermodynamic stability of the solutions obtained has to be
tested against the maximum entropy condition (or minimum of the
thermodynamic potential) over the same parameter space spanned by
the variables $h_y$, $h_z$, $v$ and $a_s$, which determine the
phase space of the present system.
However, we discuss here only the effects of crossed
Zeeman fields $h_y$, $h_z$ on the pairing temperature
$T_p$ obtained by solving the mean-field self-consistent
relations defined by Eq.~(\ref{eqn:order-parameter})
and Eq.~(\ref{eqn:number-saddle-point}) with the order parameter amplitude
set to zero, i. e., $\Delta_0 = 0$.

\begin{figure} [htb]
\includegraphics[width = 1.0\linewidth]{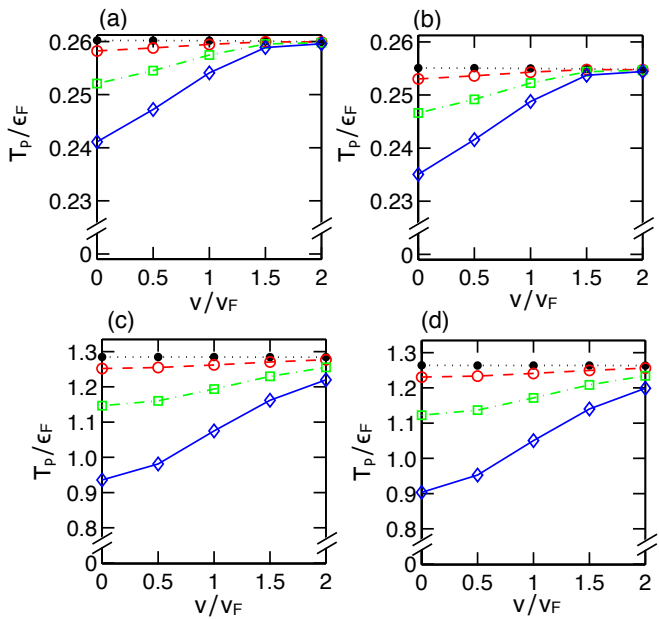}
\caption{
\label{fig:thirteen}
(color online)
Pairing temperature $T_{p}/\epsilon_F$ as a function
of ERD spin-orbit coupling
$v/v_{F}$ for selected values of $h_y$, $h_z$ and interaction parameter
$1/(k_F a_s)$. In (a) $h_y/\epsilon_{F} = 0$
and in (b) $h_y/\epsilon_{F} = 0.08$ both for $1/k_{F} a_{s} = -0.5$.
In (c) $h_y/\epsilon_{F} = 0$ and in (d) $h_y/\epsilon_{F} = 0.4$
both for $1/(k_{F} a_{s}) = 1.0$
For (a) and (b),
the black dotted line labels $h_z = 0$,
the red dashed line $h_z/\epsilon_F = 0.05$,
the green dash-dotted line $h_z/\epsilon_F = 0.1$,
and
the blue solid line $h_z/\epsilon_F = 0.15$.
In contrast, for (c) and (d),
the black dotted line labels $h_z=0$,
the red dashed line $h_z/\epsilon_F = 0.5$,
the green dash-dotted line $h_z/\epsilon_F = 1$,
and
the blue solid line $h_z/\epsilon_F = 1.5$.
Notice that it is much easier to suppress pairing in
the BCS side $(1/(k_F a_s) < 0)$, than in the BEC side
$(1/(k_F a_s) > 0)$.
}
\end{figure}

In Fig.~\ref{fig:thirteen}, we show
the pairing temperature $T_{p}/\epsilon_F$ as a function
of ERD spin-orbit coupling
$v/v_{F}$ for selected values of $h_y$, $h_z$ and interaction parameter
$1/(k_F a_s)$. The lines are guides to the eye given that the number
of points does not form a dense set.
In (a) $h_y/\epsilon_{F} = 0$
and in (b) $h_y/\epsilon_{F} = 0.08$ both at $1/(k_{F} a_{s}) = -0.5$,
showing the behavior of $T_p$ on the BCS side of unitarity.
In (c) $h_y/\epsilon_{F} = 0$ and in (d) $h_y/\epsilon_{F} = 0.4$
both at $1/(k_{F} a_{s}) = 1.0$, showing the behavior
or $T_p$ on the BEC side of unitarity.
In (a) and (b),
the black-dotted line labels $h_z = 0$,
the red-dashed line $h_z/\epsilon_F = 0.05$,
the green-dash-dotted line $h_z/\epsilon_F = 0.1$,
and
the blue-solid line $h_z/\epsilon_F = 0.15$.
However, in (c) and (d),
the black-dotted line labels $h_z=0$,
the red-dashed line $h_z/\epsilon_F = 0.5$,
the green-dash-dotted line $h_z/\epsilon_F = 1$,
and
the blue-solid line $h_z/\epsilon_F = 1.5$.
Notice that the relative suppression of pairing with zero
center of mass momentum in the BCS side $(1/(k_F a_s) < 0)$
is larger than in the BEC side $(1/(k_F a_s) > 0)$, since
it relies strongly on pairing of states only close to the
Fermi wavevectors ${\bf k}_F$ and $-{\bf k}_F$.

\begin{figure} [htb]
\includegraphics[width = 1.0\linewidth]{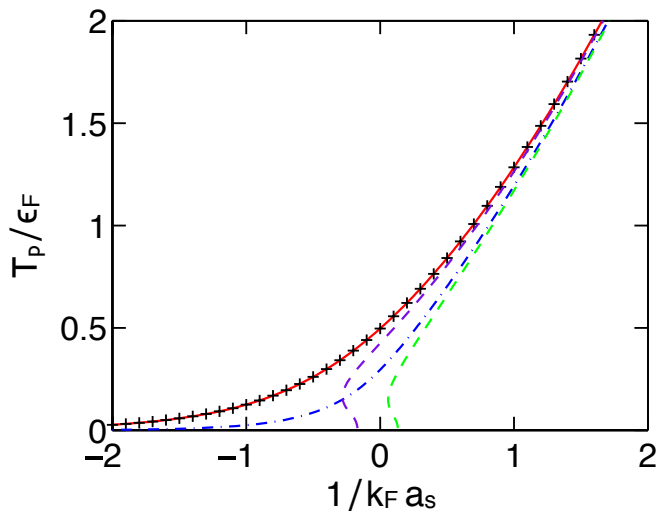}
\caption{
\label{fig:fourteen}
(color online)
The plot of the pairing temperature $T_{p}/\epsilon_F$
as a function of  $1/k_{F} a_{s}$.
The black dotted line represent the case for ERD spin-orbit coupling
$v/v_F = 0$, and Zeeman fields $h_y/\epsilon_F = h_z/\epsilon_F = 0$;
the red	solid line for corresponds
to $v/v_F = 1$, $h_y/\epsilon_F = 0$, and $h_z/\epsilon_F = 0$;
the blue dashed-dotted line
to $v/v_F = 1$, $h_y/\epsilon_F = 0$, and $h_z/\epsilon_F=1$;
the purple dashed line describes the case
of $v/v_F = 1$, $h_y/\epsilon_F = 0.4$, and $h_z = 0$;
and finally
the green dashed line corresponds to
$v/v_F = 1$, $h_y/\epsilon_F = 0.4$,
and $h_z/\epsilon_F = 1$.
}
\end{figure}

In Fig.~\ref{fig:fourteen}, we show the pairing temperature $T_p/\epsilon_F$
versus scattering parameter $1/(k_F a_s)$ for various values of the artificial
Zeeman field components $h_y$ and $h_z$ and particular values of the
ERD spin-orbit coupling $v$.
The black cross line corresponds to parameters
$v/v_F = 0$, $h_y/\epsilon_F = h_z/\epsilon_F = 0$,
where there are no ERD spin-orbit coupling and no Zeeman fields.
The red solid line corresponds to parameters
$v/v_F = 1$, $h_y/\epsilon_F = h_z/\epsilon_F = 0$.
Notice that these lines coincide, because the ERD spin-orbit field can be
gauged away producing exactly the same results for any value of
$v/v_F$ so long as $h_y/\epsilon_F = h_z/\epsilon_F = 0$.
The blue dashed-dotted line describes the case for parameters
$v/v_F = 1$, $h_y/\epsilon_F = 0$, and $h_z/\epsilon_F = 1$,
showing that the presence of Zeeman field $h_z$ in the BCS regime
produces an energy cost for pairing of Fermions with opposite momenta
and zero center-of-mass momentum, thus reducing
the pairing temperature substantially.
However, the purple dashed line describing the
situation corresponding to $v/v_F = 1$, $h_y/\epsilon_F = 0.4$, and
$h_z/\epsilon_F = 0$ shows a much stronger suppression of the
pairing temperature than for the case of
$v/v_F = 1$, $h_y/\epsilon_F = 0$, and $h_z/\epsilon_F = 1$
(blue dot-dashed line), because it becomes much more difficult
for fermions with opposite momenta to pair with
zero center-mass-momentum due to the parity violation
in the excitation spectrum of the fermions
introduced by $h_y$ when $v/v_F \ne 0$.
Lastly, the green dashed line shows the
case of $v/v_F = 1$, $h_y/\epsilon_F = 0.4$, and $h_z/\epsilon_F = 1$,
where the combined effect of the Zeeman energy cost
and parity violation lead to a dramatic reduction of the pairing
temperature in the BCS region and even near unitarity.

Another important point to emphasize in Fig.~\ref{fig:fourteen}
relates to the bending of $T_p$ in the two last cases corresponding
to the dashed purple and dashed green lines. This bending indicate
and instability of zero center-of-mass momentum pairing towards
finite center-of-mass momentum pairing, and point of infinite slope
in both curves indicates the separation of the two regimes. In the
cases where the pairing temperature reflects the critical temperature
of the system, the locations of infinite slope would correspond to
a Lifshitz point, and a small region to the left of the negative slope
regime would correspond to a superfluid with finite center of mass momentum,
which is favored due to parity violation in the helicity bands.

For the parameters discussed the pairing temperature
is not largely affected in the BEC regime, since the binding of fermions
is controlled by the emergence of two-body bound states with binding
energy $E_b$, and not by Cooper pairing in the presence of a Fermi sea.
In the BEC regime, an estimate of $T_p$ can be given by considering
the chemical equilibrium condition $\mu_B = 2\mu_F$, where $\mu_B$
is the chemical potential of the bosons formed by tightly bound fermions,
and $\mu_F$ is the chemical potential of unbound fermions.
Since $T_p/\epsilon_F \gtrsim 1$, both bosons and unbound fermions are
highly non-degenerate, and behave like classical ideal gases, in which
case the chemical potential can be directly calculated and used to obtain
the relation
$
T_p/\epsilon_F
\sim \vert E_b /\epsilon_F \vert
/
\ln \vert E_b/\epsilon_F \vert^{3/2}$ to logarithmic accuracy.
Here, the binding energy $E_b/\epsilon_F$
is a function of the interaction parameter $1/(k_F a_s)$, the ERD
spin-orbit parameter $v/v_F$ and the Zeeman fields $h_y/\epsilon_F$
and $h_z/\epsilon_F$. The logarithmic term is an entropy correction that
reduces the pairing temperature $T_p$ to a value much lower than the
absolute value of the binding energy $E_b$ of two fermions.
The pairing temperature $T_p/\epsilon_F$ is essentially the
same for values of $1/(k_F a_s) > 2$ (not shown in Fig.~\ref{fig:twelve}),
because the binding energy $\vert E_b/\epsilon_F \vert \gg
{\rm max} [ v/v_F, h_y/\epsilon_F, h_z/\epsilon_F ]$, in which
case $\vert E_b /\epsilon_F \vert \approx 2/(k_F a_s)^2$
and the pairing temperature is $T_p/\epsilon_F \sim [2/(k_F a_s)^2]/
\ln [1/(k_F a_s)]^3$, which agrees with the numerical calculations
in the regime of $1/(k_F a_s) \gg 1$.

We will not present here a discussion of the condensation or critical
temperature $T_c$ as it requires a full calculation of fluctuations effects,
which is now underway~\cite{han-seo-sdm-2012}.
However, in the extreme BEC regime, where $1/(k_F a_s) \gg 1$,
the critical temperature $T_c$ can be obtained from the Bose-Einstein
condensation temperature
$T_c \approx T_{BEC} = [n_B/\zeta(3/2)]^{2/3} 2\pi/m_B$,
where the zeta function $\zeta(3/2) \approx 2.6124$,
the density of bosons $n_B = n/2$ is half the density $n$ of fermions,
and $m_B$ is the mass of bosons which is a function of the mass
$m$ of fermions, and the parameters $v$, $h_y$ and $h_z$.
The ratio $m_B/m$ can be parameterized by the dimensionless ratios
$v/v_F$, $h_y/\epsilon_F$ and $h_z/\epsilon_F$. In the limit of
$1/(k_F a_s) \to \infty$ with finite $v/v_F$, $h_y/\epsilon_F$ and
$h_z/\epsilon_F$, the Boson mass $m_B \to 2m$, but corrections depending
on the aforementioned ratios tend to make the mass heavier, and for finite
but large $1/(k_F a_s)$ this mass increase tends to reduce $T_{BEC}$.
However, the effect of interactions between the effective bosons is very
subtle and the understanding of its dependence on the dimensionless
ratios $1/(k_F a_s)$, $v/v_F$, $h_y/\epsilon_F$, and $h_z/\epsilon_F$
is being currently investigated~\cite{han-seo-sdm-2012}.

After a detailed discussion of the effects of parity violation on
interactions, order parameter, and pairing temperature $T_p$, which
is is dramatically affected in the BCS regime, but largely
unaffected in the BEC regime for the parameter range investigated,
we are ready to present our conclusions.

\section{Conclusions}
\label{sec:conclusions}

We have analyzed the normal state of a degenerate Fermi gas in the
presence of artificial spin-orbit coupling and crossed Zeeman fields.
The specific form of the spin-orbit field chosen corresponds to a
mixture of equal Rashba and Dresselhaus (ERD) terms
${\bf h}_{ERD} ({\bf k}) = v k_x \hat{\bf y}$,
which has been experimentally realized. The artificial Zeeman field
$h_z$ along a defined spin quantization $z$ corresponds to the
Raman intensity of the laser beams, and the crossed Zeeman
field $h_y$ pointing along the same direction the ERD spin-orbit
field corresponds to the frequency detuning from the atomic
transition coupling two spin states.

In such configuration, the eigenvalues of the non-interacting problem
are obtained in a generalized helicity basis, where the helicity spin
projection points either along or opposite to the quantization axis
defined by the effective magnetic field
$
{\bf h}_{\rm eff} ({\bf k})
=
(0, h_y + v k_x, h_z).
$
In this case, we have shown that the presence of spin-orbit and
crossed Zeeman fields lead to parity violation in the excitation spectrum
of the non-interacting Fermi gas, which has immediate consequences for the
Fermi surface, spectral density and momentum distribution, which also
do not have well defined parity. Such parity violation emerges
in momentum-resolved spectroscopic quantities because they are
all functions of the magnitude $\vert {\bf h}_{\rm eff} ({\bf k})\vert$
of the effective field ${\bf h}_{\rm eff} ({\bf k}) =
(0, h_y + v k_x, h_z)$, which does not have well defined parity, since
${\bf h}_{\rm eff} (-{\bf k}) = (0, h_y - v k_x, h_z)$.

In addition, we have shown that information on parity violation
can be extracted from momentum-averaged thermodynamic properties such as
pressure, entropy, chemical potential, compressibility, and
spin-polarization as a function of crossed Zeeman field components
$h_y/\epsilon_F$ and $h_z/\epsilon_F$ for fixed spin-orbit coupling $v/v_F$
at fixed temperature $T$. A signature of parity violation is the different
behavior of any chosen thermodynamic quantity as a function of
$h_y$ for $h_z = 0$ and as a function of $h_z$ for $h_y = 0$. This anisotropy
of thermodynamic properties in the $(h_y,h_z)$ plane for finite
spin-orbit coupling $v/v_F$ is a direct reflection of the lack of
parity (inversion symmetry).

Lastly, we analyzed the effects of interactions for a degenerate Fermi
gas when parity is broken, and investigated how parity violation
influences the order parameter of a uniform superfluid and the fermion
pairing temperature. We noticed that the order parameter tensor
for a uniform superfluid in the generalized helicity basis
no longer possesses inversion symmetry, however singlet
and triplet pairing in can still be defined this basis and preserves
parity. Furthermore, we found that
the effects of parity violation are strong in the pairing temperature
of fermions, because it becomes increasingly more difficult to pair states
with zero center of mass momentum between helicity bands with
progressively larger loss of inversion symmetry.

\acknowledgements{We thank ARO (W911NF-09-1-0220) for support.}

\end{document}